\documentclass[%
 reprint,
 amsmath,amssymb,
 aps,
pra,
]{revtex4-2}

\usepackage{graphicx}% Include figure files
\usepackage{dcolumn}% Align table columns on decimal point
\usepackage{bm}% bold math
\usepackage{soul}
\usepackage{xcolor}
\usepackage{amsmath}
\DeclareMathOperator\arctanh{arctanh}
\DeclareMathOperator{\arccot}{arccot}
\usepackage{hyperref}% add hypertext capabilities
\usepackage{subcaption} % For subfigures (alternative to subfigure package)
\usepackage{booktabs} % For better table aesthetics
\usepackage{array}    % For advanced alignment options
\usepackage{float}

\newcommand{\bx}{\boldsymbol{x}}
\newcommand{\by}{\boldsymbol{y}}
\newcommand{\bp}{\boldsymbol{p}}

\begin{document}

\preprint{APS/123-QED}

\title{Electrostatic interactions between anisotropic particles}% Force line breaks with \\

\author{Harshit Joshi}
 \email{harshit.joshi@icts.res.in}
 \affiliation{
 International Centre for Theoretical Sciences, Bengaluru (ICTS-TIFR), Karnataka.}%Lines break automatically or can be forced with \\
\author{Anubhab Roy}%
\affiliation{%
 Department of Applied Mechanics, Indian Institute of Technology Madras, Chennai, Tamil Nadu
}%

\begin{abstract}
We investigate the electrostatic interactions between two charged anisotropic conductors using a combination of asymptotic and numerical methods. For widely separated particles, we employ the method of reflections to analyze the interactions. Although the formulation applies to conductors of arbitrary shapes, it is specifically implemented for spheroid-sphere systems to capture anisotropy effects in a simple configuration. In near-contact cases with axisymmetric configurations, the lubrication approximation is used to extend the analysis. Additionally, we develop a Boundary Integral Method (BIM) to study particle interactions at arbitrary separations, validating the results with asymptotic solutions for both near and far fields. We derive analytical expressions for the electrostatic force and torque on a spheroid due to another spheroid in the far-field regime.
% The electrostatic torque in this anisotropic system aligns spheroids in a broadside orientation relative to the separation vector (\hj{Comment: The broadside is for like-charged particles, for opposite charges its longside orientation. Need to change the next line a bit.}). This alignment contrasts with the behavior induced by hydrodynamic interactions, which can contribute to the instability of density fluctuations in dilute suspensions of sedimenting spheroids.
When combined with hydrodynamic effects, the electrostatic torque competes with the hydrodynamically favourable alignments of a pair of settling spheroids in certain regions while reinforcing them in others. Consequently, the inclusion of electrostatic effects may influence the instability observed in dilute suspensions of spheroids.

% \begin{description}
% Secondary publications and information retrieval purposes.
% % \item[Structure]
% % You may use the \texttt{description} environment to structure your abstract;
% % use the optional argument of the \verb+\item+ command to give the category of each item. 
% \end{description}
\end{abstract}

\maketitle

%\tableofcontents

\section{\label{sec:intro} Introduction}

Electrostatic interactions play a significant role in various natural and industrial processes, influencing behaviors across systems as diverse as atmospheric phenomena, biological assemblies, and colloidal suspensions \cite{jones1995electromechanics, wang2001electrostatic, sun2022electrostatics, edwards2014controlling, pruppacher2012microphysics}. In atmospheric science, for example, electrostatic forces are integral to cloud formation, where charged particles, including ice crystals and droplets, cluster and interact in complex ways that impact precipitation and cloud evolution \cite{pruppacher2012microphysics}. Even droplets bearing the same charge can coalesce due to electrostatic induction effects, enabling attraction through localized polarization despite net repulsion between like charges \cite{patra2022brownian,patra2023collision}. This phenomenon, while extensively studied for simple geometries like spherical particles, is less understood in realistic cases involving anisotropic interactions and irregular shapes.

One of the simplest non-spherical shapes relevant in such studies is the spheroid, a shape commonly found in atmospheric ice crystals and approximations of biological and industrial particles. To better understand the interaction of such anisotropic objects, this study focuses on the electrostatic interaction between a conducting sphere and a spheroidal body. Specifically, this work presents the first known calculation of the electrostatic torque exerted on a spheroid by a nearby sphere, which represents a key contribution to modeling how such particles align and rotate under electrostatic forces. This torque, together with the corresponding interaction forces, could be incorporated into cloud microphysics models to complement hydrodynamic models that already consider droplet interactions driven by hydrodynamic forces \cite{dhanasekaran2021collision}. In mixed-phase clouds, ice crystals collide with supercooled liquid droplets, becoming coated in a process called riming \cite{wang2000collision,naso2018collision}. Riming is a critical process in the formation of precipitation-sized hydrometeors within clouds. Precise calculation of the interaction forces between the anisotropic hydrometeor and the droplet is vital for accurately determining the collision efficiency during the riming process between ice particles and supercooled droplets.
% \\

Electrical charging mechanisms in clouds involve complex interactions between droplets, ice crystals, and graupel particles, driven by a combination of collisions and environmental factors \cite{pruppacher2012microphysics, patra2023collision}. 
Field measurements in weakly electrified clouds show that ice crystal and droplet charges are proportional to their surface areas \cite{twomey1956electrification, krasnogorskaya1969warm, colgate1970charge}.
Mechanisms such as inductive charging, which arises from the polarization of particles in an existing electric field, and convective charging, where vertical air currents separate charged particles, also play a role in cloud electrification. However, the most significant mechanism is collisional charging, where charge transfer occurs during collisions between particles. For example, when supercooled water droplets freeze upon colliding with graupel particles, charge separation occurs due to differences in ion mobility and thermal properties. In this process, smaller ice crystals typically acquire a positive charge, while graupel or hailstones gain a negative charge, with the charge separated during each collision ranging from $1\times 10^{-14}$ to $5\times 10^{-14}$ coulombs. Since collisional charging is the dominant process driving charge separation in clouds, and ice crystals are inherently anisotropic, understanding the role of particle anisotropy and their electrostatic interactions is crucial for improving our understanding of cloud electrification.

% \textbf{A para here on charging mechanisms, typical charge values for droplets and ice crystals. Check introduction of \cite{patra2023collision} and chapter 18 of \cite{pruppacher2012microphysics}}.

% \\
Analytical methods for determining electrostatic forces and torques on multiple conductors are limited to simple geometries such as sphere-sphere \cite{lekner2012electrostatics} and spheroid-spheroid in specific configurations \cite{derbenev2020electrostatic}. In this work we extend this computation to two spheroidal conductors in a generic configuration in the far field regime. The far field calculations are carried out using the method of reflections, widely used in the problems of micro-hydrodynamics \cite{kim2013microhydrodynamics}, and described in detail in the appendix \ref{app:RefMethod}. 
Having obtained the electrostatic interaction between two spheroids, we explore the role of anisotropy in the simpler, yet unexplored electrostatic interaction between a spheroid and a sphere. This system is sufficient to capture the anisotropy in the problem and provides a manageable parameter space over which relevant quantities can be analyzed. We use Boundary Integral Method (BIM) to uniformly capture the electrostatic interaction in both far and near field regimes. We compare BIM with the method of reflections to determine the proximity at which the method of reflections starts to lose accuracy for closely spaced conductors.
We derive an analytical expression for the electrostatic force and torque in the far-field regime using the first reflection, applicable to both spheroid-sphere and spheroid-spheroid systems. It is speculated that incorporating electrostatic torque in a dilute suspension of charged spheroids may modify the previously observed instability in density fluctuations of uncharged spheroids.
% The electrostatic torque in the like-charged spheroid-sphere system tends to orient the spheroid in a broadside configuration relative to the separation vector. This behavior contrasts sharply with the alignment observed under purely hydrodynamic interactions. The implications of this distinction are explored in the concluding section.

\section{Methods}

\subsection{Potential matrix formulation}

The electrostatic interaction between multiple conductors involves determining the potential on the surface of each conductor, given total charge on each conductor. This information is sufficient to determine the total electrostatic energy of the system and hence compute forces and torques on each conductor. 
The governing equation for the potential outside the conductors is simply the Laplace equation. The complexity of the problem comes from the boundary conditions that need to be satisfied at the surface of each conductor.
The linearity of governing equations of electrostatics implies a linear relationship between the total charges on each conductor and the potential on their surfaces.
The proportionality constant is called the potential matrix $\boldsymbol\Phi_M$ \cite{landau2013electrodynamics, bonnecaze1990method, zangwill2013modern, stratton2007electromagnetic, lekner2021electrostatics} which only depends on the permittivity of free space $\varepsilon_0$, size and the geometry of the conductors \footnote{The more familiar capacitance matrix is simply the inverse of the potential matrix.}. Since we are interested in two body electrostatic interaction, the connection between charges $Q_1$ and $Q_2$ and the potentials $V_1$ and $V_2$ on the surface of the conductors $S_1$ and $S_2$ is given by
\begin{equation}
\label{eq:Phimatrix}
\begin{pmatrix}
    V_1 \\
    V_2
\end{pmatrix} = 
\frac{1}{4\pi\varepsilon_0 a}
\begin{pmatrix}
    \Phi_{11} & \Phi_{12} \\
    \Phi_{21} & \Phi_{22} 
\end{pmatrix}
\begin{pmatrix}
    Q_1 \\
    Q_2
\end{pmatrix},
\end{equation}
where $a$ is the typical size of the conductors and $\Phi_{ij}$, $i,j\in \{1,2\}$, are the dimensionless elements of the potential matrix, $\boldsymbol\Phi_M$, which depends on the relative position, orientations and the geometry of the two conductors. Using the reciprocal theorem, one can show that the potential matrix is symmetric, i.e. $\boldsymbol\Phi_M^T=\boldsymbol\Phi_M$ \cite{landau2013electrodynamics, zangwill2013modern, bonnecaze1990method}.

The subsequent sections are concerned with the calculation of the potential matrix $\boldsymbol{\Phi}_M$ of a spheroid-sphere system in the far field, near field and uniformly valid regimes. Before undertaking full numerical calculations, we will first examine two distinct asymptotic limits: when the particles are widely separated and when they are nearly touching.

% Once the potential matrix $\boldsymbol\Phi_M$ is determined, the electrostatic energy $W$ of the system is simply given by
% \begin{equation}
%     \label{eq:WenergyGen}
%     W = \frac{1}{2} \boldsymbol{Q}^T\cdot \boldsymbol\Phi_M\cdot \boldsymbol{Q}\, ; \quad \boldsymbol{Q} \equiv [Q_1 \quad Q_2]^T.      
% \end{equation}

\subsection{Far field interactions: Method of reflections}

The method of reflection is an iterative approach that progressively satisfies boundary conditions on surfaces by incorporating corrections from each preceding iteration \cite{kim2013microhydrodynamics}. The solution to each iteration is given by the multipole expansions, which yields a perturbation series in $a/R$, where $a$ is typical size of the conductors and $R$ is their typical separation. 
A detailed description of this method in context of electrostatics is given in the appendix \ref{app:RefMethod}. Here, we briefly mention the common terminologies of this method. 
\begin{figure}
	\includegraphics[width=1.0\columnwidth]{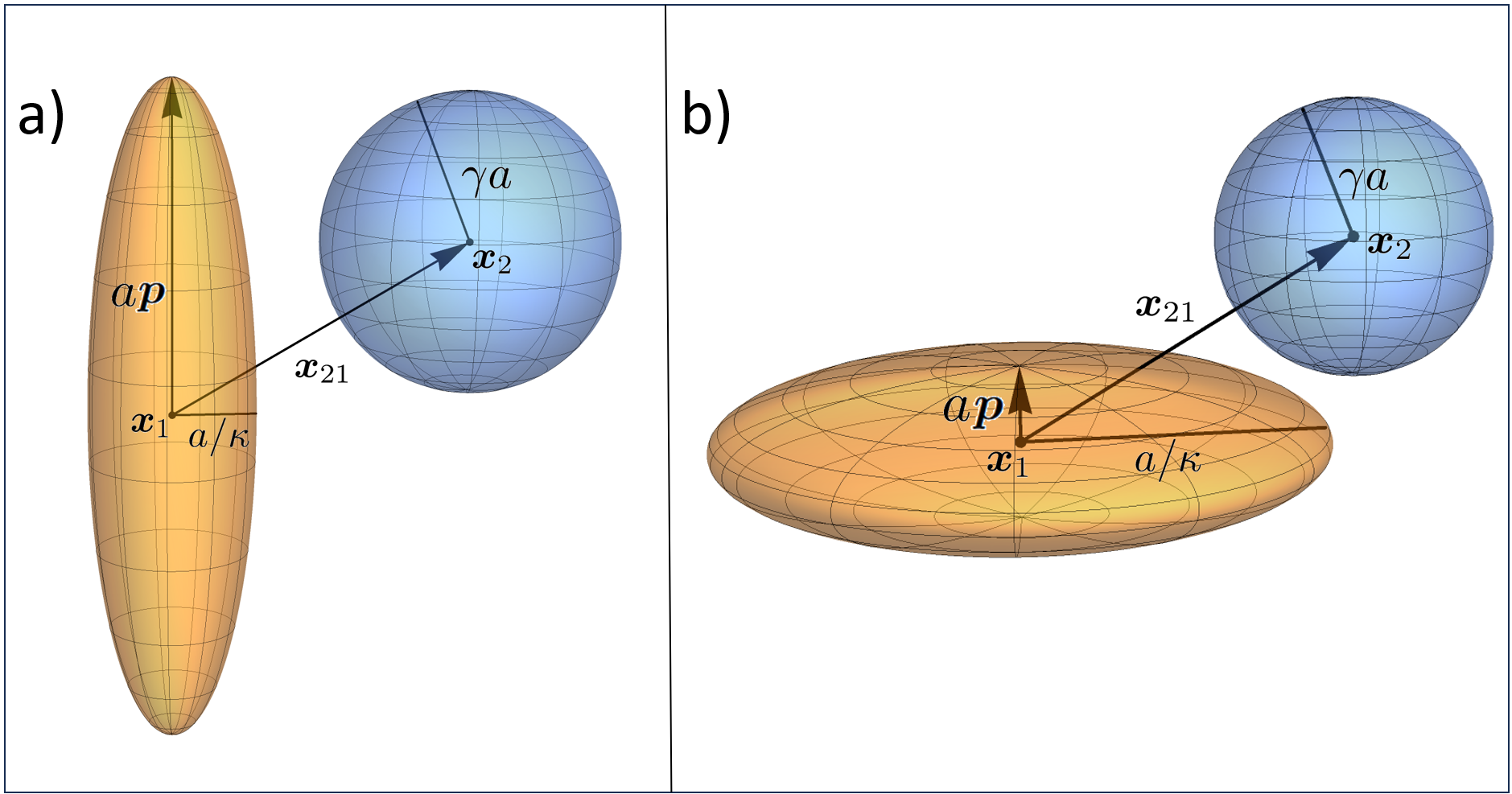}
	\caption{A schematic illustrating the geometric setup for electrostatic pair interactions between a spheroid and a sphere in a generic, non-axisymmetric configuration. The unit vector $\boldsymbol{p}$ represents the orientation of the spheroid, $a$ denoting size of the spheroid, $\kappa$ denoting its aspect ratio, $\gamma$ denoting the size ratio of sphere to spheroid.
    a) Prolate spheroid and a sphere. b) Oblate spheroid and a sphere.}
	\label{fig:notations}
\end{figure}
Consider a \textit{prolate} spheroid $S_1$, carrying a total charge $Q_1$, centered at $\bx_1$ with $a$ as the distance from its centre to the pole along the symmetry axis denoted by the unit vector $\bp$ (see figure \ref{fig:notations}). The spheroid's aspect ratio $\kappa (>1)$ is defined as the ratio of $a$ to the its equatorial radius lying perpendicular to $\bp$, and its eccentricity is $e=\sqrt{1-\kappa^{-2}}$.
The surface of this prolate spheroid is given by
\begin{equation}
    \label{eq:prolateSurface}
    (\bx-\bx_1)\cdot \left[ \frac{\bp \bp}{a^2} + \frac{(\boldsymbol{\delta}-\bp\bp)}{a^2\kappa^{-2}} \right]\cdot (\bx-\bx_1) = 1, \, \,  \bx \in S_1.
\end{equation}
The second conductor is a sphere $S_2$ centered at $\bx_2$ with radius $\gamma a$ and total charge $Q_2$, the surface of which is given by
\begin{equation}
    \label{eq:sphereSurface}
    (\bx-\bx_2)\cdot (\bx-\bx_2) = (\gamma a)^2, \quad \bx \in S_2.
\end{equation}
The relative separation vector between them is $\bx_{21}\equiv \bx_2-\bx_1\ \equiv -\bx_{12}$.
The \textit{first reflection} approximation accounts for the correction of potential fields produced by the sphere and spheroids as if they were isolated. The corresponding potential matrix in this case is accurate only upto $\mathcal{O}(a/R)$. The elements of the potential matrix for a \textit{prolate spheroid} are given by:
\begin{subequations}
    \label{eq:Phi1RefRod}
\begin{gather}
    \Phi_{11}^{(1)} = e^{-1}\arctanh e, \\
    \Phi_{12}^{(1)} = \Phi_{21}^{(1)}(\bx_{12},\bp) = \frac{1}{2e} \log\left( \frac{z_{12}-ae-R_-}{z_{12}+ae-R_+}\right), \\
    \Phi_{22}^{(1)} = \gamma^{-1}, 
\end{gather}
\end{subequations}
where
\begin{subequations}
    \label{eq:R1R2}
\begin{gather}
    R_\pm \equiv \sqrt{\rho_{12}^2 + (z_{12} \pm ae)^2},
    \\
    \rho_{12}^2 \equiv \bx_{12}\cdot(\boldsymbol{\delta}-\bp\bp)\cdot \bx_{12}, \\
    z_{12} \equiv \bx_{12}\cdot\bp. 
\end{gather}
\end{subequations}
Here we use the notation $\Phi_{ij}^{(n)}$ to represent the $ij$-th element of the potential matrix upto $n^{th}$ reflection.
Note that upto first reflection correction the effect of interaction is only captured by the the off-diagonal terms. 
Now, the \textit{second reflection} accounts for the correction in the potential fields produced in response to the first reflected fields. The corresponding potential matrix in this case is accurate upto $\mathcal{O}(a^4/R^4)$, with the elements for a \textit{prolate spheroid} given by:
\begin{widetext}
\begin{subequations}
    \label{eq:Phi1Ref2Rod}
\begin{gather}
    \Phi_{11}^{(2)}(\bx_{12},\bp) = \Phi_{11}^{(1)} -\frac{a^2\gamma^3}{4e^2}\left[ \left( \frac{1}{R_-} - \frac{1}{R_+}\right)^2 + \rho_{12}^2\left( \frac{1}{R_+(z_{12}+ae-R_+)} - \frac{1}{R_-(z_{12}-ae-R_-)}  \right)^2 \right], \\
    \Phi_{12}^{(2)}(\bx_{12},\bp) = \Phi_{21}^{(2)}(\bx_{12},\bp) = \Phi_{12}^{(1)}(\bx_{12},\bp) \\
    \Phi_{22}^{(2)}(\bx_{12},\bp) = \Phi_{22}^{(1)} - \frac{9}{4a^2 e^6}\Bigg[ 
    X^C_p\left\{ R_- - R_+ + z_{12}\log\left( \frac{z_{12}-ae-R_-}{z_{12}+ae-R_+} \right) \right\}^2 
    \nonumber \\
    + \frac{1}{4}Y^C_p
    \left\{ \frac{z_{12}}{\rho_{12}}(R_- - R_+) + \frac{ae}{\rho_{12}}(R_- + R_+)  - \rho_{12}\log\left( \frac{z_{12}-ae-R_-}{z_{12}+ae-R_+} \right) \right\}^2 \Bigg],
\end{gather}
\end{subequations}
\end{widetext}
where 
\begin{subequations}
    \label{eq:XcYcProlate}
\begin{gather}
    X^C_p \equiv \frac{e^3}{3}(\arctanh e-e)^{-1},\\ 
    Y^C_p \equiv \frac{2e^3}{3}\left( \frac{e}{1-e^2}-\arctanh e \right)^{-1}. 
\end{gather}
\end{subequations}

Now consider an \textit{oblate} spheroid $S_1$ centered at $\bx_1$ with $a$ as the distance from its centre to the pole along the symmetry axis denoted by the unit vector $\bp$ (see figure \ref{fig:notations}). Its aspect ratio is $\kappa (<1)$, with an eccentricity of $e=\sqrt{1-\kappa^2}$ and it carries a total charge $Q_1$. The surface of this oblate spheroid $S_1$ is again given by \eqref{eq:prolateSurface} with the only difference being $\kappa<1$.
The second conductor $S_2$ is again a sphere of radius $\gamma a$, centered at $\bx_2$, carrying a total charge $Q_2$.
To obtain the corresponding potential matrix of the spheroid-sphere system we use the eccentricity transformation $e\to \dfrac{i e}{\sqrt{1-e^2}}$ on the corresponding expressions of the prolate spheroid \cite{shatz2004singularity}. Therefore, for an oblate spheroid and a sphere, we have
\begin{subequations}
    \label{eq:Phi1RefDisk}
\begin{gather}
    \Phi_{11}^{(1)} = \frac{\kappa \arcsin e}{e}, \\
    \Phi_{12}^{(1)} = \Phi_{21}^{(1)}(\bx_{12},\bp) = \frac{\kappa}{e} \arccot\left( \frac{z_{12}-u}{v-ae/\kappa}\right), \\
    \Phi_{22}^{(1)} = \gamma^{-1},
\end{gather}
\end{subequations}
% \hj{Note that $\gamma_{BIM} = \gamma \kappa$}.
where $z_{12}$ is given by equation \eqref{eq:R1R2} and $u$ and $v$ are given by
\begin{subequations}
    \label{eq:uvEqn}
\begin{gather}
    u \equiv \sqrt{ \frac{\mu}{2} +
    \sqrt{\frac{\mu^2}{4} + \frac{a^2e^2}{\kappa^2}z_{12}^2} }\, ; \, \, \, \mu \equiv |\bx_{12}|^2 - \frac{a^2e^2}{\kappa^2} ,\\ 
    v \equiv \frac{a e z_{12}}{\kappa u}.
\end{gather}
\end{subequations}
Similarly, the second reflection corrections are given by
\begin{widetext}
\begin{subequations}
    \label{eq:Phi1Ref2Disk}
\begin{gather}
    \Phi_{11}^{(2)}(\bx_{12},\bp) = \Phi_{11}^{(1)} -\frac{\kappa^2a^2\gamma^3}{4 e^2}\left[ \left( \frac{2 v}{u^2+v^2} \right)^2 + \rho_{12}^2\left\{ \frac{4 ae\kappa^{-1} z_{12}-2(z_{12}v+ae\kappa^{-1} u)}{(u^2+v^2)((z_{12}-u)^2+(ae\kappa^{-1}-v)^2)}  \right\}^2 \right], \\
    \Phi_{12}^{(2)}(\bx_{12},\bp) = \Phi_{21}^{(2)}(\bx_{12},\bp) = \Phi_{12}^{(1)}(\bx_{12},\bp) \\
    \Phi_{22}^{(2)}(\bx_{12},\bp) = \Phi_{22}^{(1)} - \frac{9 \kappa^6}{a^2 e^6}\Bigg[ 
    X^C_o\left\{ v - z_{12}\arccot\left( \frac{z_{12}-u}{v-ae\kappa^{-1}} \right) \right\}^2 
    \nonumber \\
    + \frac{1}{4}Y^C_o
    \left\{ \frac{ae\kappa^{-1} u - z_{12} v}{\rho_{12}} - \rho_{12}\arccot\left( \frac{z_{12}-u}{v-ae\kappa^{-1}} \right) \right\}^2 \Bigg],
\end{gather}
\end{subequations}
\end{widetext}
where 
\begin{subequations}
\begin{gather}
    \label{eq:XcYcOblate}
    X^C_o \equiv \frac{e^3}{3}[e(1-e^2) - (1-e^2)^{3/2}\arcsin e]^{-1},\\ 
    Y^C_o \equiv \frac{2e^3}{3}\left[ e(1-e^2)^2 - (1-e^2)^{3/2}\arcsin e \right]^{-1}. 
\end{gather}
\end{subequations}

The potential matrix for two spherical conductors can be obtained by taking the limit $e\to 0$ in the potential matrix expression for a prolate spheroid. Therefore, for a spherical conductor $S_1$ of radius $a$, centered at $\bx_1$ and another spherical conductor $S_2$ of radius $\gamma a$, centered at $\bx_2$, the elements of the potential matrix upto the second reflection are given by:
\begin{subequations}
    \label{eq:Phi2RefSpheres}
\begin{gather}
    \Phi_{11}^{(2)}(|\bx_{21}|) = 1-\frac{\gamma^3 a^4}{|\bx_{21}|^4}, \\
    \Phi_{12}^{(2)}(|\bx_{21}|) = \Phi_{21}^{(2)}(|\bx_{21}|) = \frac{1}{|\bx_{21}|}, \\
    \Phi_{22}^{(1)}(|\bx_{21}|) = \frac{1}{\gamma}-\frac{a^4}{|\bx_{21}|^4}.
\end{gather}
\end{subequations}

\subsection{Near contact interaction: Lubrication approximation}

% The near contact asymptotic form of the electrostatic interactions can be obtained using the lubrication appr 

Using the lubrication approximation for the spheroid-sphere system in the axisymmetric configuration involves solving the Laplace equation for the potential field $\phi(\bx)$ near the gap of thickness $a\epsilon$ between the conductors. 
% \hj{Be careful for oblate spheroid!}
Using polar coordinates with $z$ coordinate along the symmetry axis $\bp$ and $r$ coordinate transverse to $\bp$, the boundary value problem to be solved is
\begin{subequations}    
    \label{eq:LaplaceLub}
    \begin{gather}
    \nabla^2\phi = \frac{\partial^2 \phi}{\partial z^2} + \frac{1}{r}\frac{\partial}{\partial     
    r}\left(r\frac{\partial \phi}{\partial r} \right) = 0, \\
    \phi = 
    \begin{cases}
        V_1 & z = h_1(r), \\
        V_2 & z = h_2(r),
    \end{cases}
    \end{gather}
\end{subequations}
The surface of the spheroid and the sphere can be expanded as
\begin{subequations}
    \label{eq:hEqns}
    \begin{gather}
    \frac{h_1(r)}{a\epsilon} = 1 + \frac{\kappa^2 r^2}{2\epsilon a^2} + \frac{1}{8}\frac{\kappa^4 r^4}{\epsilon a^4} + \mathcal{O}\left(\frac{\kappa^6 r^6}{\epsilon a^6}\right), \\
    \frac{h_2(r)}{a\epsilon} = -\frac{1}{2}\frac{r^2}{\epsilon\gamma a^2} - \frac{1}{8}\frac{r^4}{\epsilon \gamma^3 a^4} + \mathcal{O}\left(\frac{r^6}{\epsilon \gamma^5 a^6}\right).
    \end{gather} 
\end{subequations}
Defining the stretched coordinates $R\equiv r/(a\sqrt{\epsilon})$ and $Z \equiv z/(a\epsilon)$, we have
\begin{subequations}
    \label{eq:hNDEqns}
    \begin{gather}
    H_1(R) = 1 + \frac{\kappa^2 R^2}{2} + \frac{\epsilon \kappa^4 R^4}{8} + \mathcal{O}(\epsilon^2), \\
    H_2(R) = -\frac{R^2}{2\gamma} - \frac{\epsilon R^4}{8\gamma^3} + \mathcal{O}(\epsilon^2).
    \end{gather} 
\end{subequations}
Rewriting the Laplace equation in terms of the stretched coordinates, we have
\begin{subequations}
    \label{eq:LaplaceStch}
\begin{gather}
    \frac{\partial^2 \phi}{\partial Z^2} + \frac{\epsilon}{R} \frac{\partial}{\partial R}\left( R\frac{\partial \phi}{\partial R} \right) = 0, \\
    \phi = 
    \begin{cases}
        V_1 & Z = H_1(R), \\
        V_2 & Z = H_2(R),
    \end{cases}
    \end{gather}
\end{subequations}
The solution can be expanded in the perturbation series as $\phi = \phi_0 + \epsilon \phi_1 + \mathcal{O}(\epsilon^2)$.
The total charge $Q_1$ and $Q_2$ on the spheroid $S_1$ and sphere $S_2$ are given by
\begin{equation}
    \label{eq:QLub}
    Q_\alpha = -\varepsilon_0 \oint_{S_\alpha} \boldsymbol{\nabla}\phi\cdot\hat{\boldsymbol{n}}_\alpha \, dS_\alpha, \quad \alpha\in\{1,2 \},
\end{equation}
where $\hat{\boldsymbol{n}}_\alpha$ represents the unit normal pointing out of the surface $S_\alpha$.
The electrostatic force $\boldsymbol{F}_1$ on the spheroid is given by
\begin{equation}
    \label{eq:F1Lub}
    \boldsymbol{F}_1 = \frac{\varepsilon_0}{2}\oint_{S_1} \left| \boldsymbol{\nabla}\phi\cdot\hat{\boldsymbol{n}}_1 \right|^2 \hat{\boldsymbol{n}}_1\, dS_1.  
\end{equation}
Using the zeroth order solution $\phi_0$, the charge difference $\Delta Q_{12}=Q_1-Q_2$ is given by:
\begin{equation}
    \label{eq:Q1Lub2}
    \Delta Q_{12} = \frac{4\pi a \varepsilon_0 \gamma \Delta V_{12}}{1+\gamma\kappa^2}\left[\log\left( \frac{1+\gamma\kappa^2}{2\gamma\kappa \epsilon} \right) + \delta \right] + \mathcal{O}(\epsilon),
\end{equation}
where $\Delta V_{12} \equiv V_1-V_2$ and $\delta$ is an $\mathcal{O}(1)$ constant which has to be determined using the outer solution. The weak logarithmic singularity is insufficient to overpower the $\delta$ correction, even at very small separations $\epsilon$, and therefore $\delta$ cannot be neglected. 
The forces $\boldsymbol{F}_1$ and $\boldsymbol{F}_2$ are given by
\begin{equation}
    \label{eq:F1Lub2}
    \boldsymbol{F}_1 = -\boldsymbol{F}_2\sim \frac{-\hat{\bx}_{12}(1+\gamma \kappa^2) \Delta Q_{12}^2}{16\pi a^2 \varepsilon_0 \gamma \epsilon \left[ \log\left( \frac{1+\gamma \kappa^2}{2\gamma\kappa\epsilon} \right) + \delta \right]^2 } .
\end{equation}
Note that for unequal total charges $\Delta Q_{12}\neq 0$, the electrostatic forces at close range are attractive, regardless of whether the conductors carry like or unlike charges. The force expression \eqref{eq:F1Lub2} reduces to the near contact force between two spheres for $\kappa = 1$ (\cite{khair2013electrostatic, lekner2012electrostatics}).

We rewrite equation \eqref{eq:Q1Lub2} in terms of $\delta$ as
\begin{equation}
    \label{eq:deltaEqn}
    \delta = \lim_{\epsilon\to 0} \left\{\frac{(1+\gamma\kappa^2)\Delta Q_{12}}{4\pi a\varepsilon_0 \gamma \Delta V_{12}} - \log\left( \frac{1+\gamma\kappa^2}{2\gamma\kappa \epsilon} \right)\right\}.
\end{equation}
We shall use Boundary Integral Method (BIM) to evaluate the right hand side of the above equation for $\epsilon\ll 1$ and thus obtain $\delta$. 
The numerical values of the RHS of equation \eqref{eq:deltaEqn} will have small variations with $\epsilon$ even when $\epsilon\ll1$. This is due to the fact that the numerical errors in BIM increases as the surfaces approach each other \cite{bagge2023accurate,prosperetti2009computational}. 
The error in the numerical measurement of $\delta$, i.e. $\Delta \delta$, gives error on the forces $|\Delta F|$ (see equation \eqref{eq:F1Lub2}) as
\begin{equation}
    \label{eq:relErrorF}
    |\Delta F| = |\boldsymbol{F}_1|\left|\frac{2\Delta \delta} {\left[\log\left( \frac{1+\gamma \kappa^2}{2\gamma\kappa\epsilon} \right) + \delta\right] }\right|.
\end{equation} 
Note that the relative error in the forces decreases with $\epsilon$.
Once the $\delta$ is obtained, the lubrication force \eqref{eq:F1Lub2} gives electrostatic forces in the configurations where minimum separation between the conductors become vanishingly small. 

\subsection{Boundary Integral Method}

The method of reflections is primarily effective for far-field interactions. Achieving higher accuracy requires additional reflections, but each successive reflection adds significant complexity in the analytical expressions. To compute the interactions in both far and near field regimes numerically, we use the Boundary Integral Method (BIM). The BIM formulation is well established for various linear partial differential equations, including the Laplace equation \cite{pozrikidis2002practical, martinsson2019fast, gorman2024electrostatic}. A brief formulation of the BIM for the electrostatic problem with total charges specified on each conductors is given in the appendix \ref{app:bim}. Here we outline the main integral equations to be solved numerically to compute the potential matrix for a spheroid $S_1$ (both prolate and oblate) and a sphere $S_2$. The potentials on the surface of the conductors are given by:
\begin{gather}
    \label{eq:VBIM}
    \varepsilon_0 V_\alpha = \frac{1}{|S_\alpha|}\oint_{S_\alpha} q_\alpha(\bx)\, dS_\alpha(\bx); \quad \alpha \in \{1,2 \},
\end{gather}    
where $|S_\alpha|$ is the surface area of the conductor $S_\alpha$. The fields $q_\alpha$ are obtained by solving the second kind integral equation on every point $\bx_{s\alpha}$ on the surface of conductor $S_\alpha$:
\begin{multline}
\label{eq:intEqn}
    \begin{bmatrix}
        \mathcal{L}^d_{11} + \mathcal{P}^c_{11} + \mathbb{I} & \mathcal{L}^d_{12} \\
        \mathcal{L}^d_{21} & \mathcal{L}^d_{22} + \mathcal{P}^c_{22} + \mathbb{I} 
    \end{bmatrix} \begin{bmatrix}
        q_1 \\
        q_2
    \end{bmatrix} \\
    = 
    \begin{bmatrix}
        Q_1\mathcal{G}(\bx_{s1}, \bx_1) + Q_2\mathcal{G}(\bx_{s1}, \bx_2) \\
        Q_1\mathcal{G}(\bx_{s2}, \bx_1) + Q_2\mathcal{G}(\bx_{s2}, \bx_2)
    \end{bmatrix},
\end{multline}
where $Q_1,\, Q_2$ are the charges on the conductors $S_1$ and $S_2$, respectively, and $\mathcal{G}$ is the Greens function of the Laplacian, given by
\begin{equation}
    \label{eq:LaplaceGreen}
    \mathcal{G}(\bx, \bx_0) \equiv \frac{1}{4\pi |\bx-\bx_0|}.
\end{equation}
The integral operators are defined as:
\begin{subequations}
    \label{eq:intOperators}
\begin{gather}
    \mathcal{L}^d_{\alpha\beta}q_\beta(\bx_s) \equiv 2\oint_{S_\beta} q_\beta(\bx)\boldsymbol{\hat n}_\beta \cdot \boldsymbol{\nabla}_{\bx}\mathcal{G}(\bx,\bx_s)\, dS_\beta(\bx), \\
    \mathcal{P}_{\alpha\beta}^c q_\beta \equiv \frac{1}{|S_\alpha|} \delta_{\alpha\beta}\oint_{S_\beta} q_\beta(\bx)\, dS_\beta(\bx); \quad \bx_s \in S_\alpha,
\end{gather}
\end{subequations}
$\alpha,\, \beta \in \{1,2 \}$.
The equations \eqref{eq:VBIM} and \eqref{eq:intEqn} are used to determine the potential matrix. The integral equation \eqref{eq:intEqn} is solved using GMRES iterations \cite{saad1986gmres, bagge2023accurate} and the integrals on the surfaces are evaluated using the Gaussian quadrature \cite{golub1969calculation, bagge2023accurate}.

\subsection{Electrostatic force and Torque}

When particles carry an electric charge, they can experience strong mutual interactions. Precisely calculating the electric forces and torques acting on these charged particles is crucial across a wide range of physical systems, including biological cells, ice crystals, and granular materials. These force calculations are essential for predicting particle dynamics, such as their trajectories and the potential for aggregation. The electrostatic force and torque on each conductor can be computed by taking derivatives of the electrostatic energy of the system. The electrostatic energy of the spheroid-sphere system is given by:
\begin{equation}
    \label{eq:Energy}
    W(|\bx_{21}|, \hat{\bx}_{21}\cdot \bp) = \frac{1}{2} \boldsymbol{Q}^T\cdot \boldsymbol{\Phi}_M(|\bx_{21}|, \hat{\bx}_{21}\cdot \bp) \cdot \boldsymbol{Q},
\end{equation}
with $\boldsymbol{Q}\equiv [Q_1 \quad Q_2]^T$, where the spheroid centered at $\bx_1$ carries a total charge $Q_1$ and the sphere centered at $\bx_2$ carries a total charge $Q_2$. Here $\hat{\bx}_{21}$ is a unit vector along the separation vector $\bx_{21} = \bx_2-\bx_1$.
The differential change in the electrostatic energy upon differential change in the relative configuration is given by
\begin{equation}
    \label{eq:dWEqn}
    dW = d\bx_{21}\cdot \boldsymbol{\nabla}_{21}W + d\bp\cdot \boldsymbol{\nabla}_{p}W.
\end{equation}
The first term in equation \eqref{eq:dWEqn} represents the negative of the work done by the electrostatic force on the sphere, $\boldsymbol{F}_2$, in moving the sphere by an amount $d\bx_{21}$. Equivalently, it represents the negative of the work done by the electrostatic force on the spheroid, $\boldsymbol{F}_1$, in moving the spheroid by an amount $-d\bx_{21}$. Therefore, the electrostatic forces on the conductors are given by:
\begin{equation}
    \label{eq:Feqn}
    \boldsymbol{F}_1 = -\boldsymbol{F}_2 = \boldsymbol{\nabla}_{\bx_{21}}W(|\bx_{21}|, \hat{\bx}_{21}\cdot \bp).
\end{equation}
The second term shows that there is energy expense in changing the orientation of the spheroid. This shows that the electrostatic force on the spheroid does not act at its centre. Thus, an electrostatic torque $\boldsymbol{T}_1$ acts on the spheroid about its centre. The work done by the electrostatic force on the spheroid in changing its orientation can be written in terms of $\boldsymbol{T}_1$ as $\boldsymbol{T}_1\cdot \hat{\boldsymbol{n}}\, d\theta$, where $\hat{\boldsymbol{n}}$ is the axis about which $\bp$ is rotated by an angle $d\theta$, i.e. $d\bp = d\theta\, \hat{\boldsymbol{n}}\times \bp$. Equating this to the second term in equation \eqref{eq:dWEqn} gives the torque on the spheroid about its centre as
\begin{equation}
    \label{eq:T1Eq}
    \boldsymbol{T}_1 = -\bp \times\boldsymbol{\nabla}_{\bp}W(|\bx_{21}|, \hat{\bx}_{21}\cdot \bp).
\end{equation}
The change in configuration due to the change in the orientation vector $\bp = d\theta\, \hat{\boldsymbol{n}}\times \bp$ is equivalent to keeping the spheroid's orientation fixed but rotating the separation vector $\bx_{21}$ about the spheroid's centre, the opposite way, such that $d\bx_{21} = -d\theta\, \hat{\boldsymbol{n}}\times \bx_{21}$. The work done on the sphere by $\boldsymbol{F}_2$ in this case is simply, $\boldsymbol{F}_2 \cdot (-d\theta\, \hat{\boldsymbol{n}}\times \bx_{21}) = -(\bx_{21}\times \boldsymbol{F}_2 )\cdot \hat{\boldsymbol{n}}\, d\theta \equiv \boldsymbol{T}_2 \cdot (-\hat{\boldsymbol{n}}d\theta)$. This shows the torque $\boldsymbol{T}_2$ on the sphere is simply
\begin{equation}
    \label{eq:T2Eq}
    \boldsymbol{T}_2 = \bx_{21}\times \boldsymbol{F}_2.
\end{equation}
It is easy to see using equations \eqref{eq:Feqn}, \eqref{eq:T1Eq} and \eqref{eq:T2Eq} that $\boldsymbol{T}_1 = -\boldsymbol{T}_2$, and hence the total angular momentum of the system is conserved.

\section{Results}

The parameter space to be explored contains the aspect ratio of spheroid $\kappa$ and ratio of the sphere's radius to the spheroid's semi-major axis $\gamma$ for various configurations given by $\bx_{21}$ and $\bp$. For a given $\kappa$, we fix the value of $\gamma$ such that the surface area of the spheroid is same as that of the sphere. We look at three different aspect ratios $\kappa \in \{1, 4, 0.25\}$. The first case corresponds to the electrostatic interaction between two identical spheres, results of which are well known \cite{lekner2012electrostatics}. This serves as a benchmark for our general results for spheroid-sphere interactions. The other two cases corresponds to a prolate and an oblate spheroid, respectively.  

% \begin{center}
% \begin{tabular}{ |c|c|c|c| } 
%  \hline
%  $\kappa$ & 1 & 4 & 0.25  \\ 
%  \hline
%  $\gamma$ & 1 & 0.445 & 3.01 \\ 
%  \hline
% \end{tabular}
% \end{center}
\begin{table}[h!]
\centering
\renewcommand{\arraystretch}{1.2} % Adjust row height
\begin{tabular}{@{}>{\centering\arraybackslash}p{2cm} >{\centering\arraybackslash}p{2cm} >{\centering\arraybackslash}p{2cm} >{\centering\arraybackslash}p{2cm}@{}}
\toprule
$\kappa$ & 1 & 4 & 0.25 \\ 
\midrule
$\gamma$ & 1 & 0.445 & 3.01 \\ 
\bottomrule
\end{tabular}
\caption{The values of $\kappa$ and $\gamma$ used in the numerical calculations for the three systems—sphere-sphere, prolate spheroid-sphere, and oblate spheroid-sphere-are provided.}
\label{tab:kappa_beta}
\end{table}

% \hj{Add statement about non-dimensionalization of quantities to be plotted.}

\subsection{Elements of the potential matrix}

\begin{figure}
	\includegraphics[width=1.0\columnwidth]{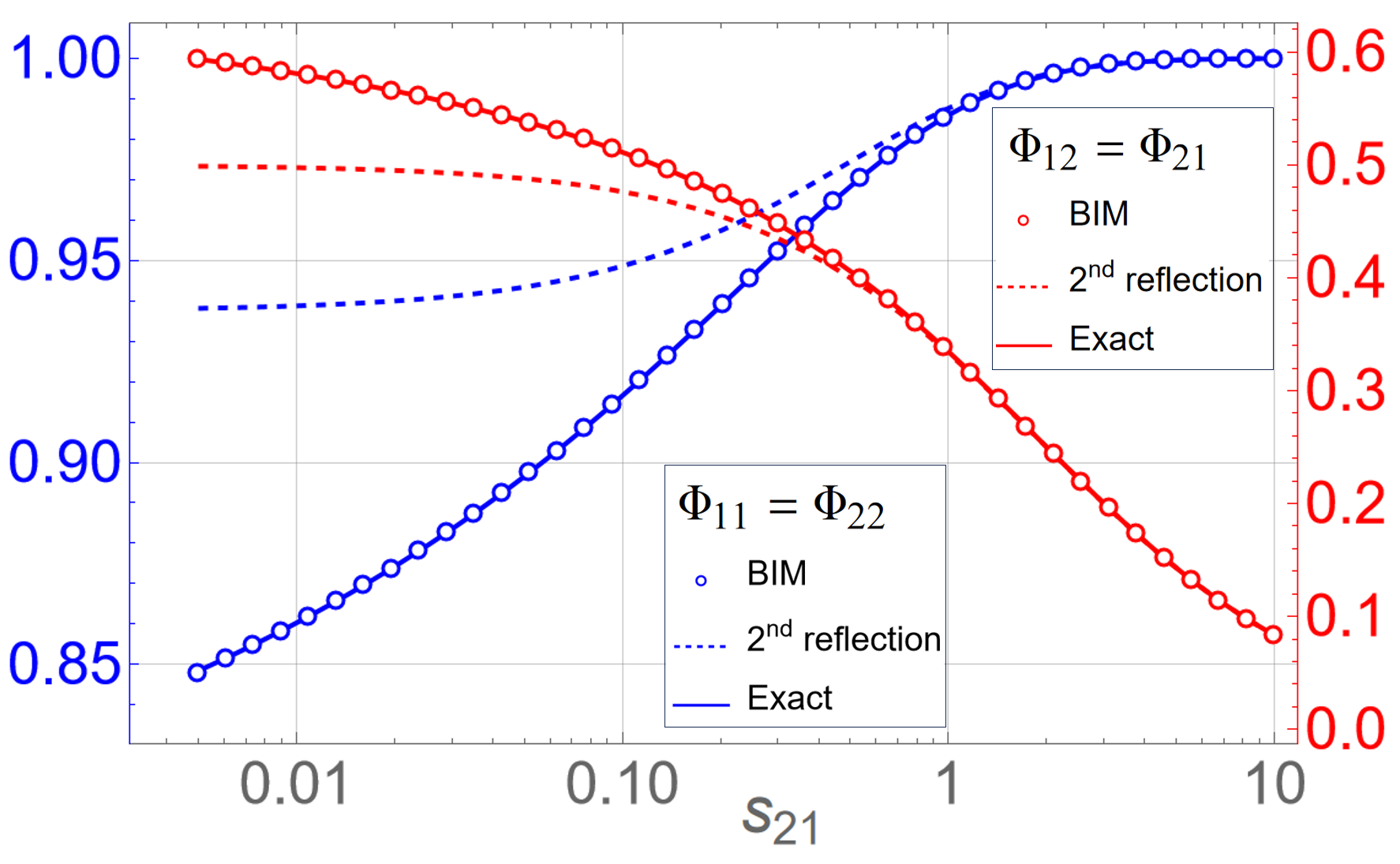}
	\caption{Elements of the potential matrix $\Phi_M$ (see \eqref{eq:Phimatrix}) as a function of dimensionless minimum separation between the two spheres, $s_{21}=|\bx_{21}|/a - 2$. The second reflection is decent upto the separations of the order of the size of the spheres. The exact result in terms of an infinite series can be found in \cite{lekner2012electrostatics}.}
	\label{fig:phiSpheres}
\end{figure}

The elements of the potential matrix are defined in equation \eqref{eq:Phimatrix}. For the case of two spheres ($\kappa=1$), the exact expression is known from Lekner \cite{lekner2012electrostatics} and the second reflection results are given by equation \eqref{eq:Phi2RefSpheres}. The comparison between second reflection, BIM and exact expression shows that the second reflection performs well down to minimum separations between spheres comparable to their size, see figure \ref{fig:phiSpheres}. This also validates both the second reflection and the BIM.

For the case of electrostatic interactions between a spheroid and a sphere, the exact expressions of the potential matrix are not known to the best of our knowledge. The potential matrix depends on both the separation between the conductors $|\bx_{21}|$ and the relative configuration of the conductors $\cos(\psi) \equiv \hat{\bx}_{21}\cdot \bp$. The minimum separation between the centres of the conductors when they are just touching depends on $\psi$ and is denoted by $d_{\text{min}}(\psi)$. This minimum separation can be determined numerically by finding the roots $\bx^*$ (point of contact) and $d^*$ of the following equations
\begin{subequations}
    \label{eq:dMinFinding}
    \begin{gather}
    \left|\frac{\boldsymbol{n}_1}{|\boldsymbol{n}_1|} + \frac{
    \boldsymbol{n}_2}{|\boldsymbol{n}_2|}\right| = 0, \label{eq:cndtn1}\\
    \left|\bx^*+\gamma a \frac{\boldsymbol{n}_1}{|\boldsymbol{n}_1|} - d^*\hat{\bx}_{21}\right| = 0, \label{eq:cndtn2} \\
    |\boldsymbol{n}_2|^2 = \gamma^2a^2 \label{eq:cndtn3} \\
    \bx^*\cdot (\bp \times \hat{\bx}_{21}) = 0 \label{eq:cndtn4},
    \end{gather}
\end{subequations}
where $\boldsymbol{n}_1$ and $\boldsymbol{n}_2$ are the (non-normalized) normal vectors to the spheroid and sphere at $\bx^*$, given by:
\begin{subequations}
    \label{eq:normalVec}
    \begin{gather}
    \boldsymbol{n}_1 \equiv \left[\frac{\bp\bp}{a^2} + \frac{(\boldsymbol{\delta}-\bp\bp)}{a^2\kappa^{-2}}\right]\cdot(\bx^*-\bx_1), \\\boldsymbol{n}_2 \equiv \bx^*-\bx_1 - d^*\hat{\bx}_{21}.
    \end{gather}
\end{subequations}
The four equations \eqref{eq:dMinFinding} uniquely determines $\bx^*$ and $d^* = d_{\text{min}}(\psi)$.
Note that $\hat{\bx}_{21}$ is given by a unit vector making an angle $\psi$ with $\bp$, which doesn't require specifying $d^*$. 

\begin{figure}
	\includegraphics[width=1.0\columnwidth]{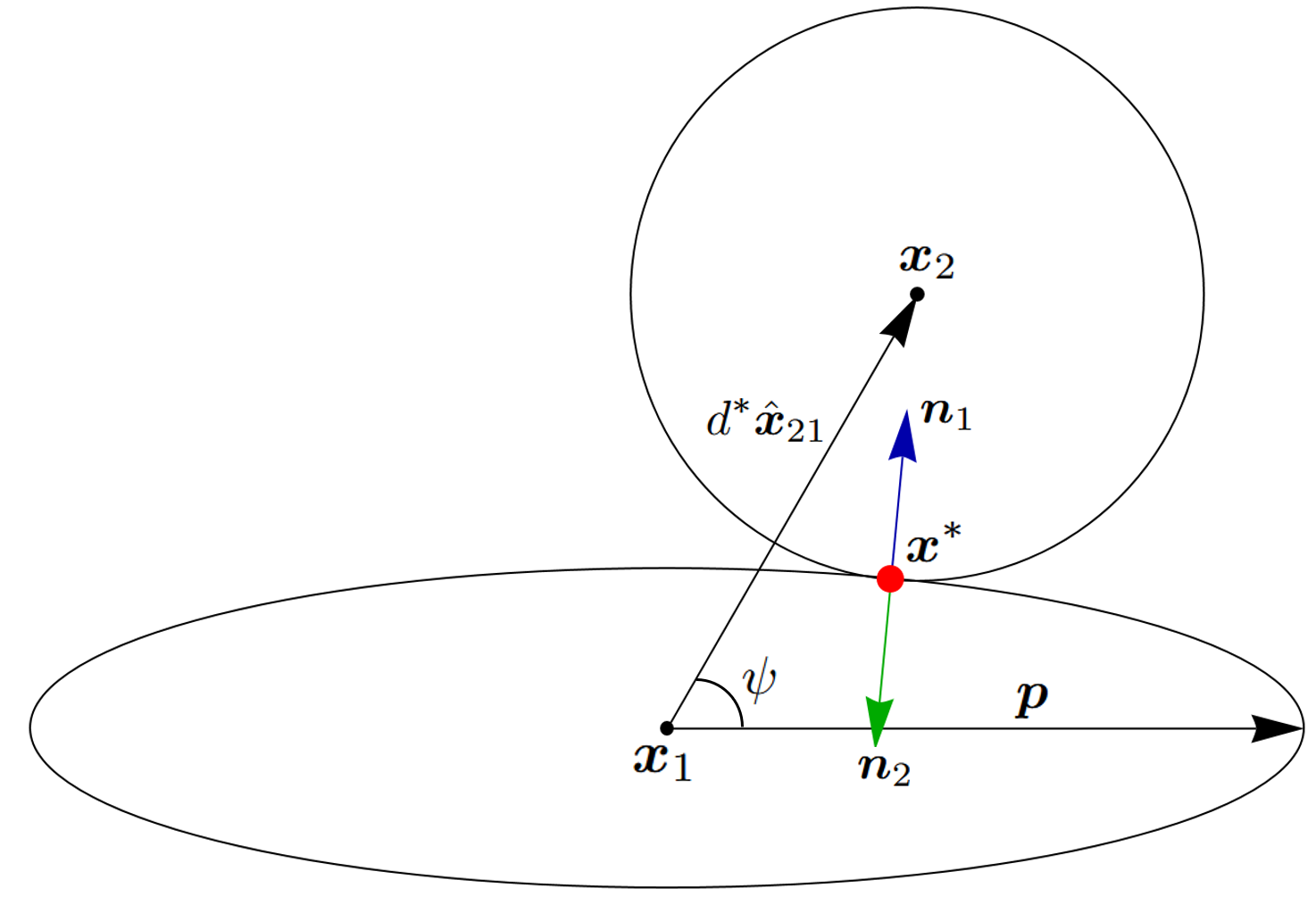}
	\caption{A schematic showing the point of contact $\bx^*$, the minimum distance $d^*=d_{\text{min}}(\psi)$ and other relevant quantities for the case of a prolate spheroid and a sphere. The relative sizes of the conductors are proportional to their respective scales.}
	\label{fig:schematic}
\end{figure}

A schematic representing $\bx^*$, $d^*$ and other relevant quantities is shown in figure \ref{fig:schematic}.

The physical interpretation of equation \eqref{eq:dMinFinding} is as follows:
\begin{enumerate}
    \item Equation \eqref{eq:cndtn1} enforces that the normals of the sphere and the spheroid are oriented anti-parallel to each other.
    \item Equation \eqref{eq:cndtn2} ensures that $\bx^*$ is the point of contact.
    \item Equation \eqref{eq:cndtn3} ensures that $\bx^*$ lies at the surface of the sphere.
    \item Equation \eqref{eq:cndtn4} ensures that $\bx^*$ lies in the plane defined by $\bp$ and $\hat{\bx}_{21}$.
\end{enumerate}
% where $\bx_2 = \bx_1 + d_{\min} \hat{\bx}_{21}$. 

\begin{figure}
	\includegraphics[width=1.0\columnwidth]{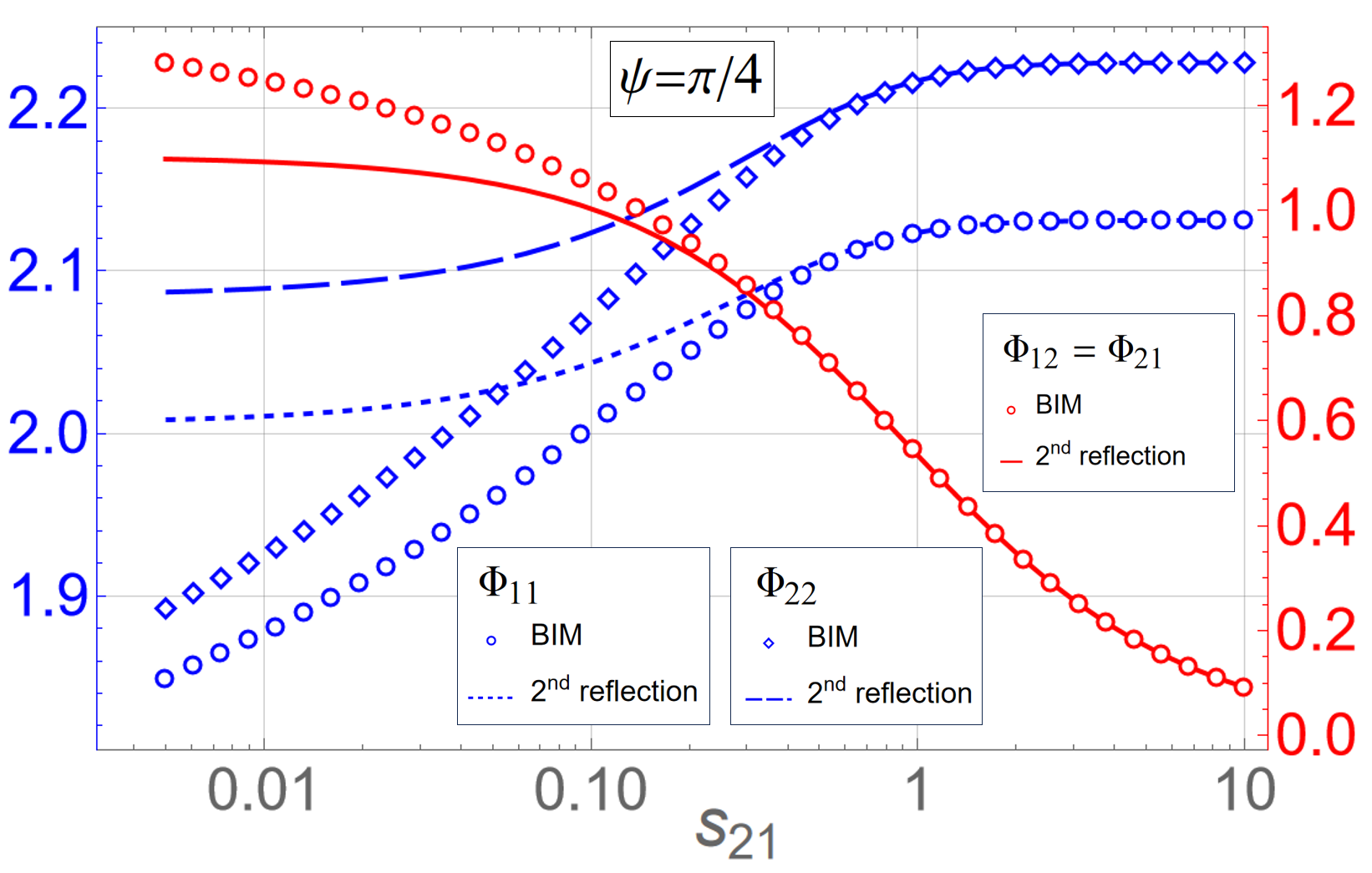}
	\caption{Elements of the potential matrix $\Phi_M$ (see \eqref{eq:Phimatrix}) as a function of dimensionless separation between the a prolate spheroid and a sphere, $s_{21}=(|\bx_{21}| - d_{\text{min}}(\psi))/a$. Here $\psi\equiv \arccos(\hat{\bx}_{21}\cdot \bp)$ and $d_{\text{min}}$ is the dimensionless center-to-center distance between the prolate spheroid and the sphere when they are just in contact.}
	\label{fig:phiProlate}
\end{figure}

For the case of a prolate spheroid and a sphere ($\kappa=4$), the second reflection results are given in equation \eqref{eq:Phi1Ref2Rod}. 
Figure \ref{fig:phiProlate} shows the elements of the potential matrix for a fixed $\psi=\pi/4$ as a function of dimensionless separation $s_{21} = (|\bx_{21}| - d_{\text{min}}(\psi))/a$. The second reflection is reliable upto $s_{21} \sim 1$.

Similarly, for the case of an oblate spheroid and a sphere ($\kappa=0.25$), the second reflection results (equation \eqref{eq:Phi1Ref2Disk}) are reliable upto $s_{21} \sim 4$. This early deviation of the second reflection method from BIM arises because the length scale used for $s_{21}$ does not correspond to the larger dimension of the oblate spheroid, specifically the equatorial radius of the oblate spheroid $a\kappa^{-1}$.

\begin{figure}
	\includegraphics[width=1.0\columnwidth]{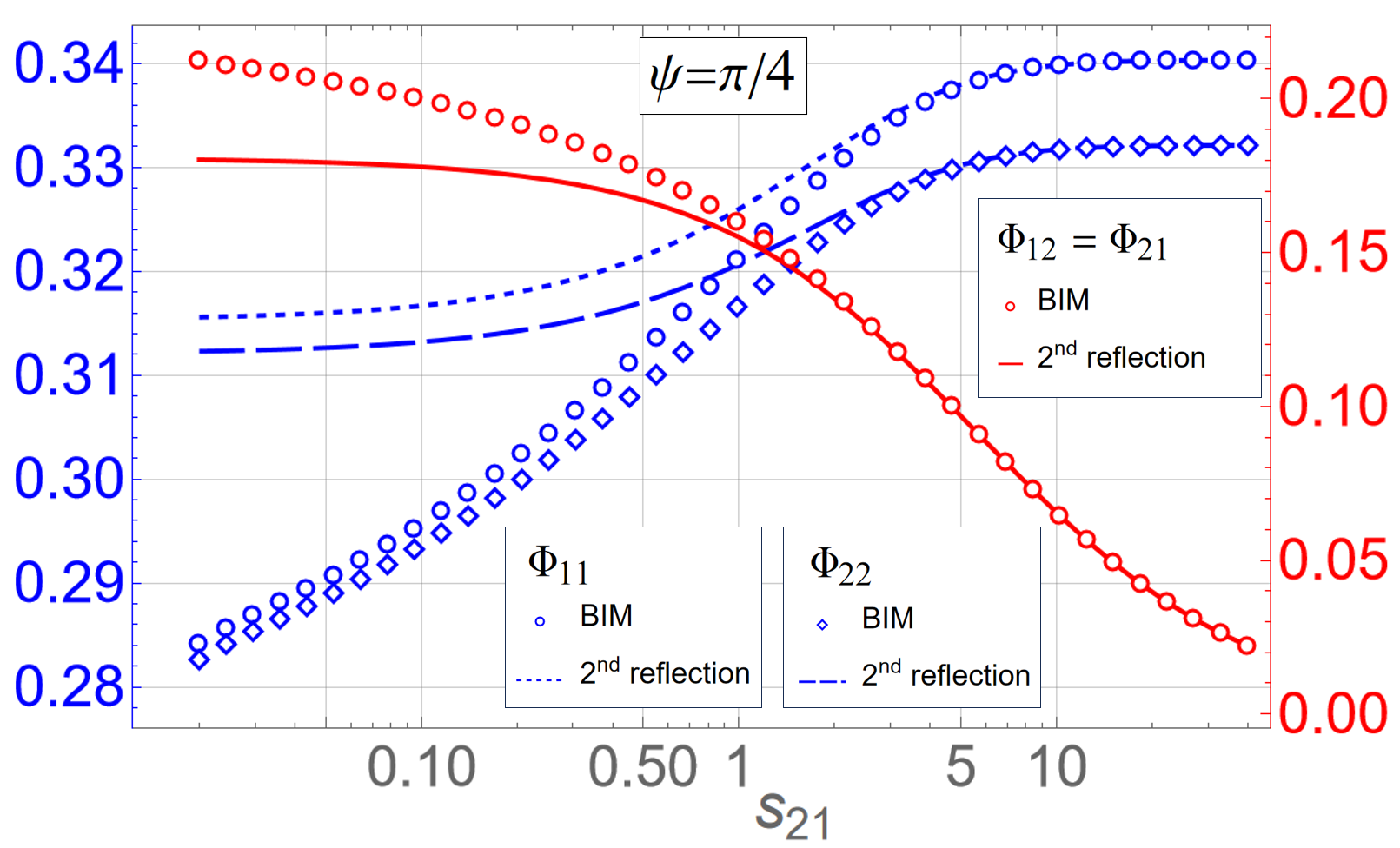}
	\caption{Elements of the potential matrix $\Phi_M$ (see \eqref{eq:Phimatrix}) as a function of dimensionless separation between the an oblate spheroid and a sphere, $s_{21}=(|\bx_{21}| - d_{\text{min}}(\psi))/a$. Here $\psi\equiv \arccos(\hat{\bx}_{21}\cdot \bp)$ and $d_{\text{min}}$ is the dimensionless center-to-center distance between the oblate spheroid and the sphere when they are just in contact.}
	\label{fig:phiOblate}
\end{figure}

\subsection{Electrostatic force}

\begin{figure}
	\includegraphics[width=1.0\columnwidth]{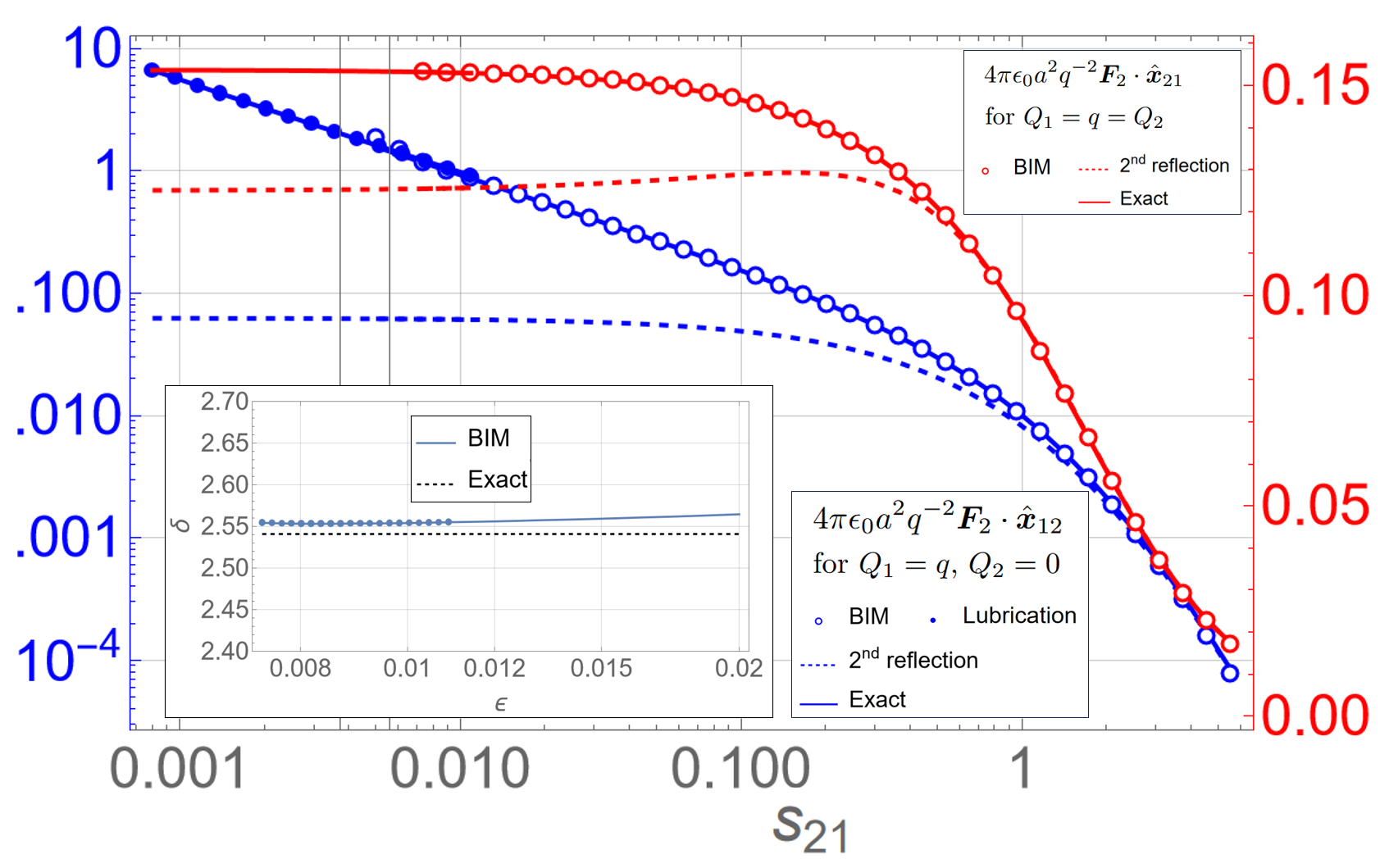}
	\caption{Dimensionless force on the second sphere as a function of dimensionless minimum separation between
    the two spheres, $s_{21} = |\bx_{21}|/a - 2$. 
    Note that the force is attractive in the case of unequal charges ($\boldsymbol{F}_2\cdot \bx_{12} > 0$).
    The filled dots are obtained using the lubrication approximation (see equation \eqref{eq:F1Lub2}) with $\delta$ obtained using BIM through equation \eqref{eq:deltaEqn}. The inset shows $\delta$ as a function of $\epsilon$, with the dots indicating the range of values over which $\delta$ is averaged to approximate it as a constant.}
	\label{fig:forceSpheres}
\end{figure}

Equation \eqref{eq:Feqn} is used to obtain the electrostatic force between the pair of conductors. This relies on differentiating the electrostatic energy obtained using the potential matrix. The exact results are available for the sphere-sphere case by \cite{lekner2012electrostatics}. The second reflection is again reliable upto $s_{21} \sim 1$. For very small separation $s_{21} \ll 1$, the BIM needs large number of collocation points on the surfaces of the conductors to converge to the solution accurately. Lubrication approximation (equation \eqref{eq:F1Lub2}) has been used for $s_{21} \ll 1$, shown by the filled dots in figure \ref{fig:forceSpheres}, with the $\delta$ fitted using the BIM results. 

\begin{figure}
	\includegraphics[width=1.0\columnwidth]{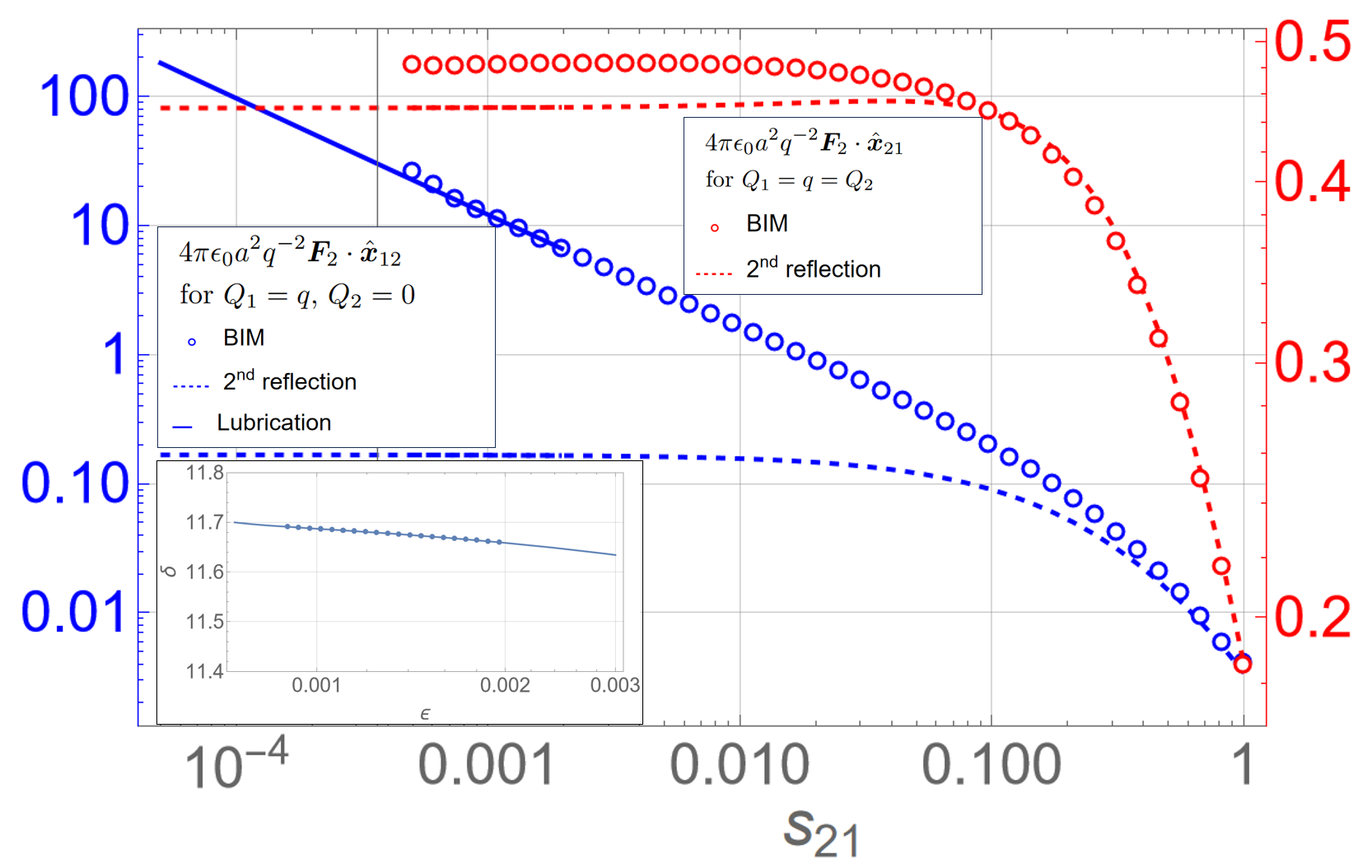}
	\caption{Dimensionless force on the second sphere as a function of dimensionless separation between
    the prolate spheroid and the sphere, $s_{21} = |\bx_{21}|/a - (1+\gamma)$, in the axisymmetric configuration ($\bp=\hat{\bx}_{21}$). 
    Note that the force is attractive in the case of unequal charges ($\boldsymbol{F}_2\cdot \bx_{12} > 0$).
    The lubrication approximation is obtained using equation \eqref{eq:F1Lub2} with $\delta$ obtained using BIM through equation \eqref{eq:deltaEqn}. The inset shows $\delta$ as a function of $\epsilon$, with the dots indicating the range of values over which $\delta$ is averaged to approximate it as a constant.}
	\label{fig:forceLubProlate}
\end{figure}

\begin{figure}
	\includegraphics[width=1.0\columnwidth]{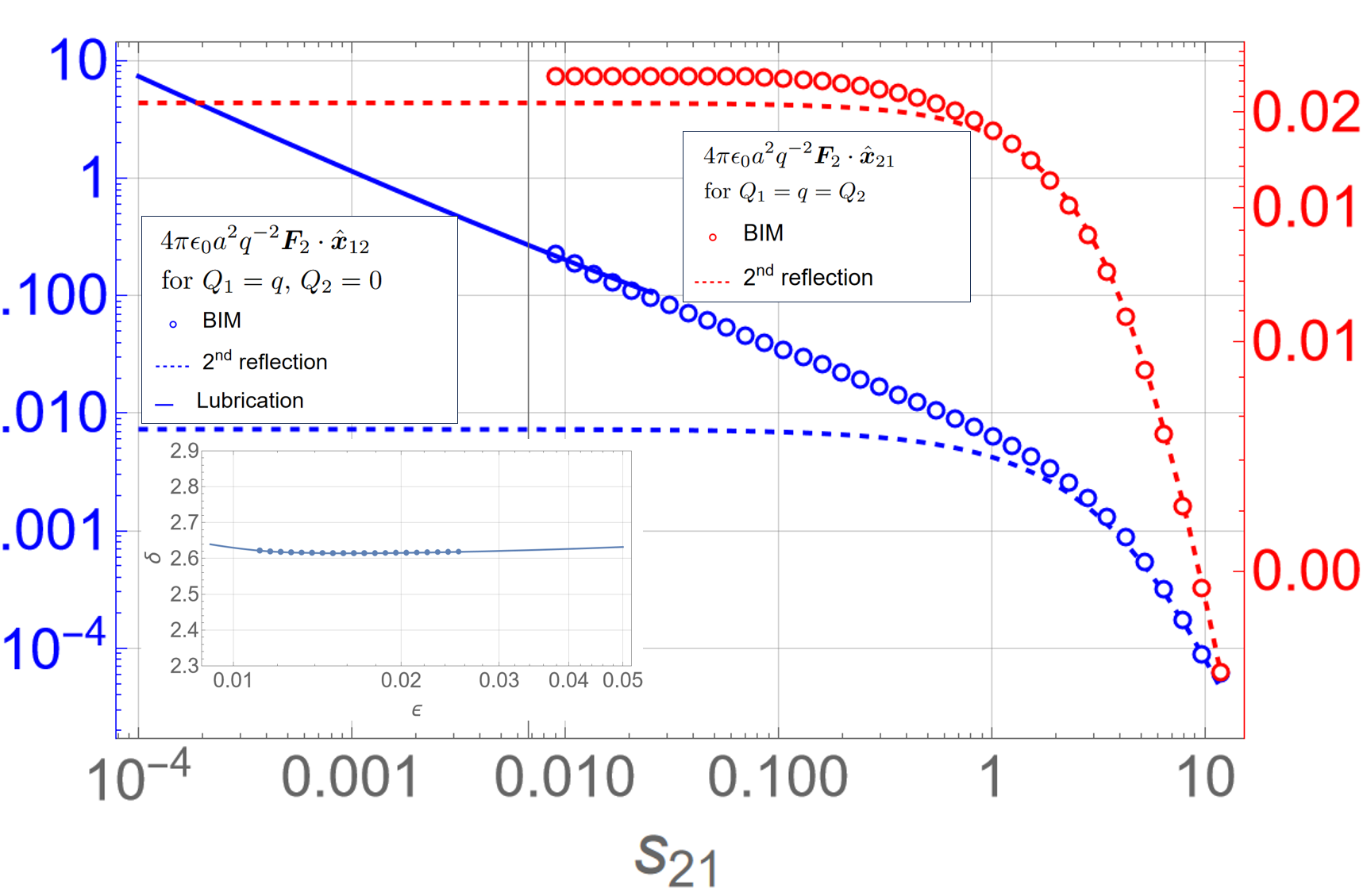}
	\caption{Dimensionless force on the second sphere as a function of dimensionless separation between
    the oblate spheroid and the sphere, $s_{21} = |\bx_{21}|/a - (1+\gamma)$, in the axisymmetric configuration ($\bp=\hat{\bx}_{21}$). 
    Note that the force is attractive in the case of unequal charges ($\boldsymbol{F}_2\cdot \bx_{12} > 0$).
    The lubrication approximation is obtained using equation \eqref{eq:F1Lub2} with $\delta$ obtained using BIM through equation \eqref{eq:deltaEqn}. The inset shows $\delta$ as a function of $\epsilon$, with the dots indicating the range of values over which $\delta$ is averaged to approximate it as a constant.}
	\label{fig:forceLubOblate}
\end{figure}

The force acting on the sphere in the axisymmetric configuration  ($\bp \cdot \hat{\bx}_{21}$) involving a prolate spheroid and a spherical conductor is shown in figure \ref{fig:forceLubProlate}. The lubrication force is given by equation \eqref{eq:F1Lub2} with the $\delta$ fitted using the BIM results. Figure \ref{fig:forceLubOblate} shows the corresponding force for the case of an oblate spheroid and a sphere.
Note that the electrostatic forces are attractive in the near contact case for unequal charges and grows unboundedly.

\begin{figure*}[t] % The '*' makes it span both columns
    \centering
    \begin{subfigure}[b]{0.24\textwidth} % 1/4 of the text width minus some spacing
        \includegraphics[width=\textwidth]{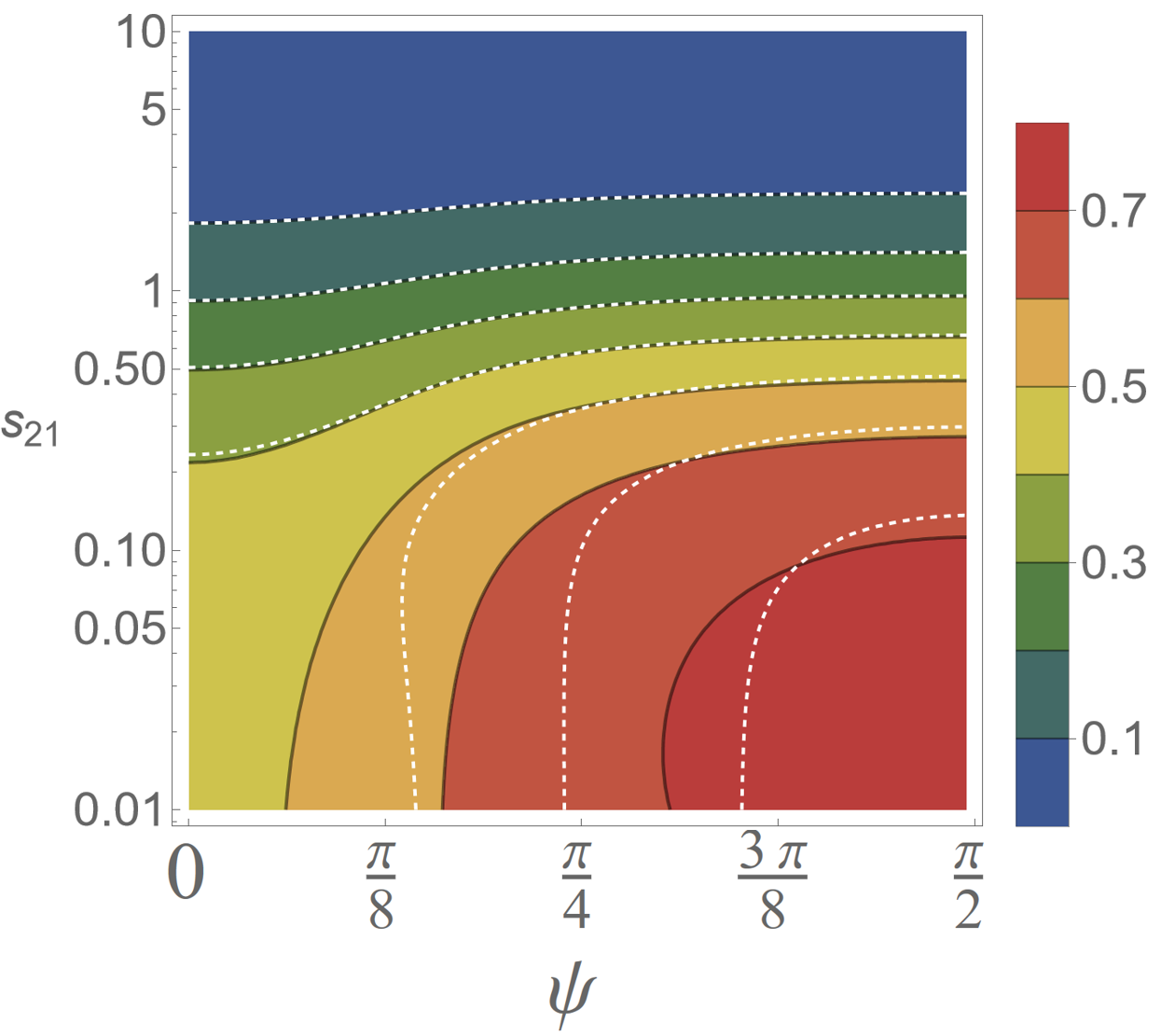}
        \caption{The prolate spheroid has charge $Q_1=q$ and the sphere has charge $Q_2=q$.}
        \label{fig:F2alongProlate11}
    \end{subfigure}
    \hfill
    \begin{subfigure}[b]{0.24\textwidth}
        \includegraphics[width=\textwidth]{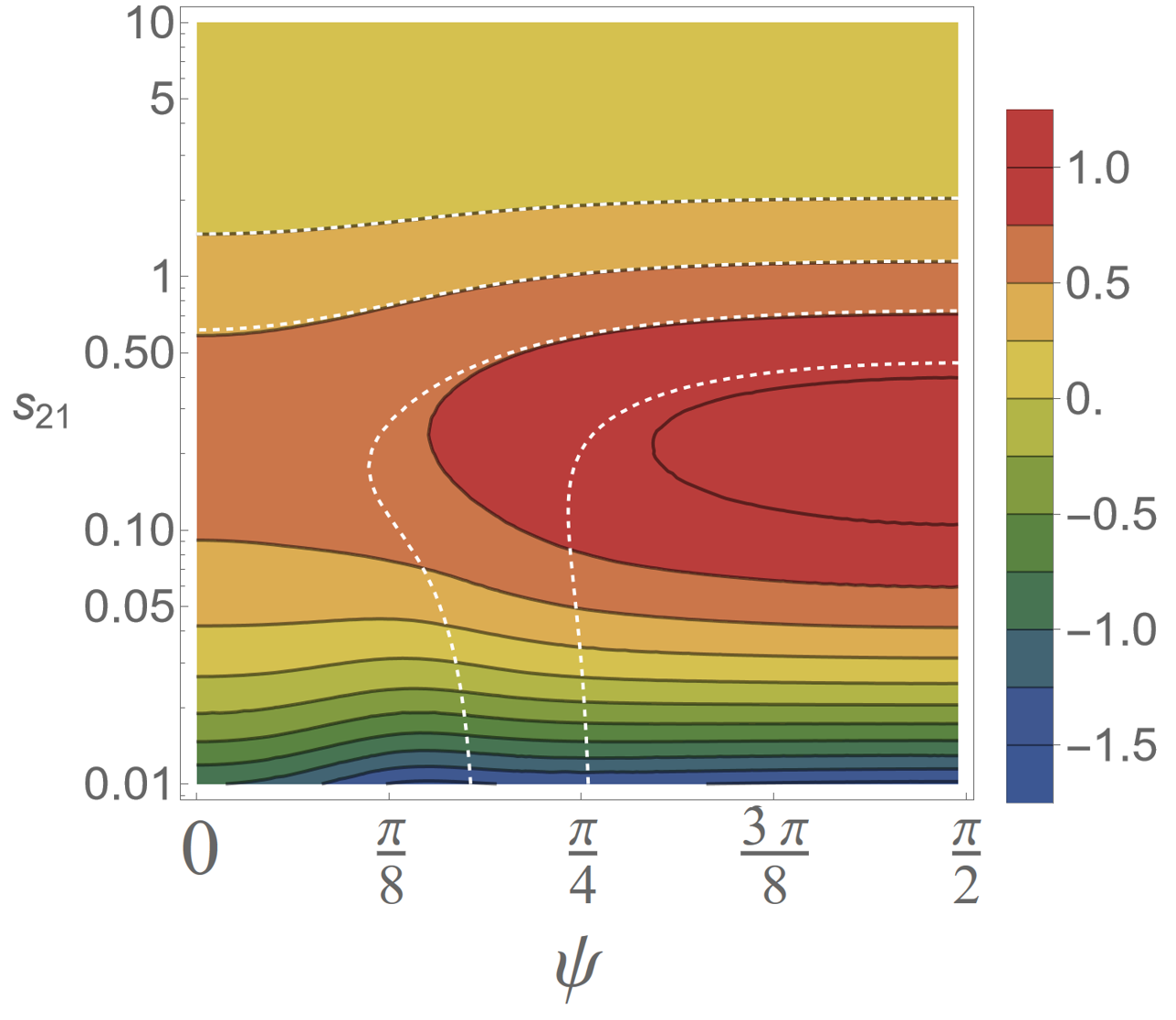}
        \caption{The prolate spheroid has charge $Q_1=q$ and the sphere has charge $Q_2=2q$.}
        \label{fig:F2alongProlate12}
    \end{subfigure}
    \hfill
    \begin{subfigure}[b]{0.24\textwidth}
        \includegraphics[width=\textwidth]{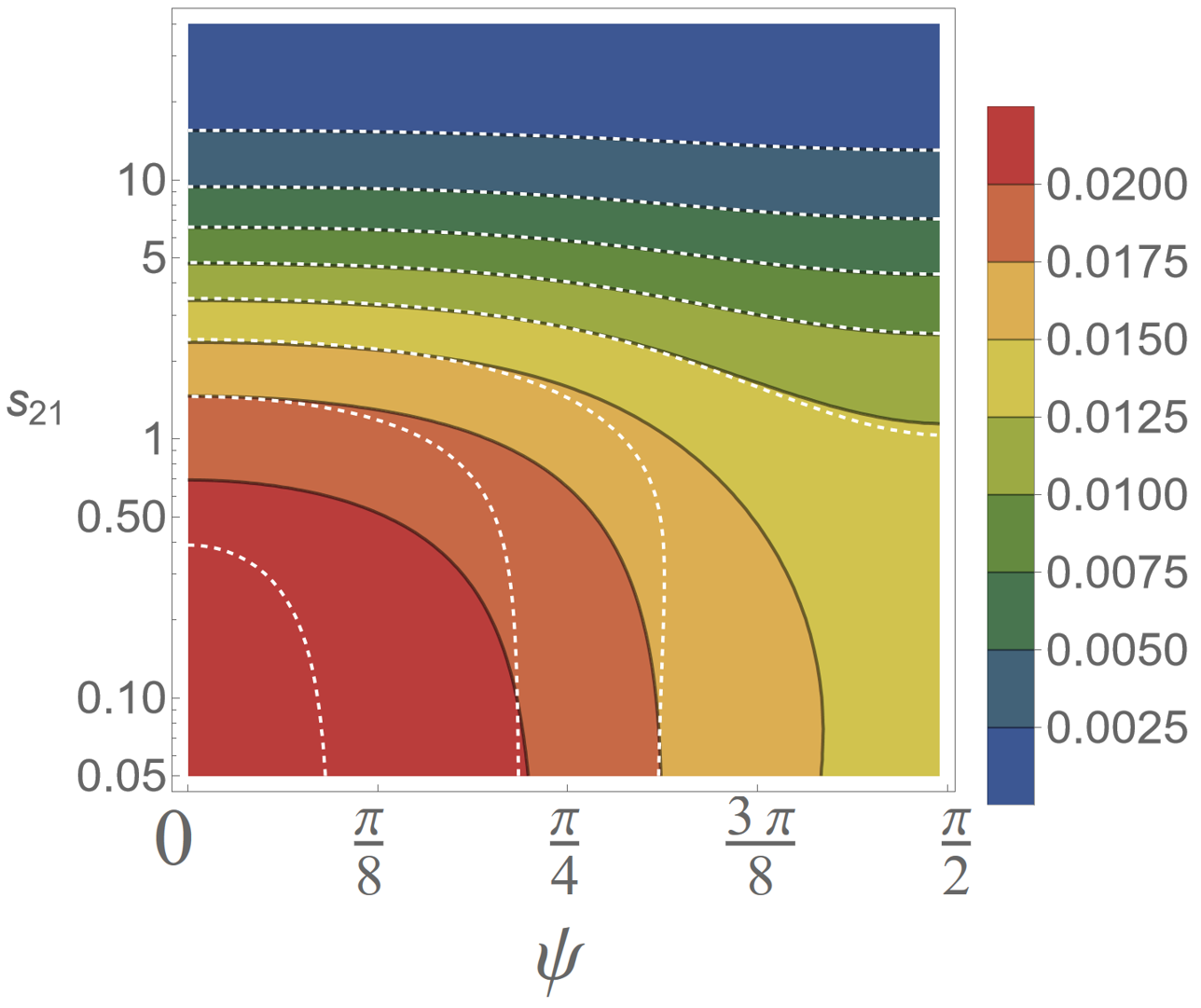}
        \caption{The oblate spheroid has charge $Q_1=q$ and the sphere has charge $Q_2=q$.}
        \label{fig:F2alongOblate11}
    \end{subfigure}
    \hfill
    \begin{subfigure}[b]{0.24\textwidth}
        \includegraphics[width=\textwidth]{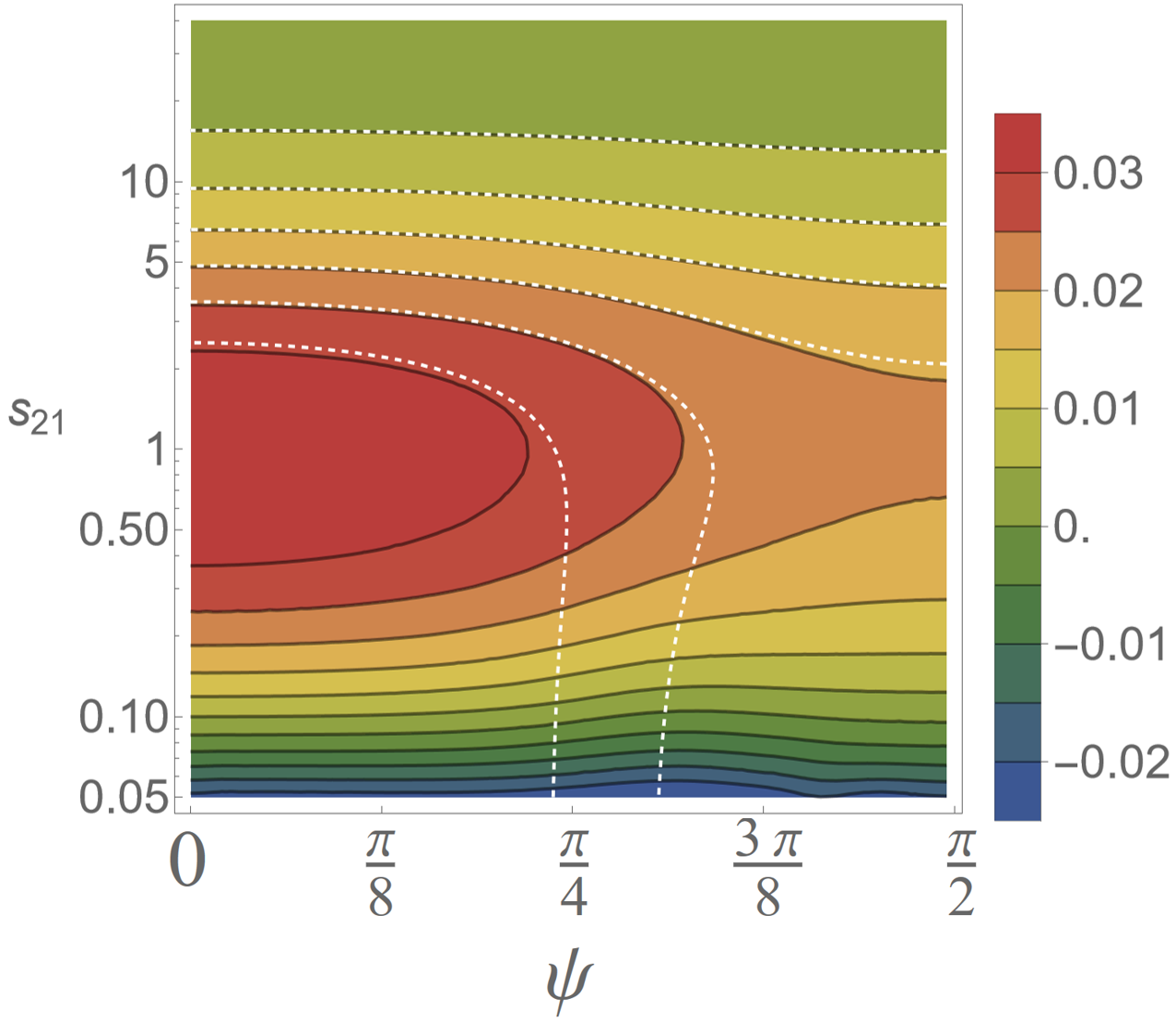}
        \caption{The oblate spheroid has charge $Q_1=q$ and the sphere has charge $Q_2=2q$.}
        \label{fig:F2alongOblate12}
    \end{subfigure}
    \caption{Contour plot of dimensionless force along the separation vector, $4\pi\varepsilon_0 a^2 q^{-2}\boldsymbol{F}_2\cdot \hat{\bx}_{21}$, as a function of $\psi\equiv \arccos(\hat{\bx}_{21}\cdot \bp)$ and $s_{21}=(|\bx_{21}| - d_{\text{min}}(\psi))/a$. The white dotted lines are the due to the second reflections.}
    \label{fig:forcesAlong}
\end{figure*}

Figure \ref{fig:forcesAlong} shows the variation of electrostatic force as a function of dimensionless separation $s_{21}$ and the relative configuration $\psi$. This captures the effect of anisotropy of the problem. 

One is often interested in the dilute regime where particle separations are much larger than their size. In this regime, the first reflection is sufficient to capture the electrostatic force. Using equation \eqref{eq:Feqn} and \eqref{eq:Phi1RefRod}, the electrostatic force for the prolate spheroid and sphere system is given by
\begin{multline}
    \label{eq:FR1}
    \boldsymbol{F}_2 \sim \frac{Q_1 Q_2}{8\pi\varepsilon_0 |\bx_{21}|}\Bigg\{ \left(\frac{1}{R_+}+\frac{1}{R_-} \right)\hat{\bx}_{21} + \frac{|\bx_{21}|}{ae} \\
    \left( \frac{1 - ae/R_+}{R_+-ae-z_{12}} - \frac{1 + ae/R_-}{R_- + ae-z_{12}} \right)(\boldsymbol{\delta} - \hat{\bx}_{21}\hat{\bx}_{21})\cdot\bp \Bigg\}
\end{multline}
where $R_-,\, R_+$ and $z_{12}$ are given by equation \eqref{eq:R1R2}. 
The corresponding electrostatic force due to the first reflection for the oblate spheroid and sphere system is given by
\begin{multline}
    \label{eq:FR1oblate}
    \boldsymbol{F}_2 \sim \frac{Q_1Q_2}{4\pi\varepsilon_0 |\bx_{12}|}\Bigg\{
    \frac{a^2e^2z_{12}^2 + \kappa^2 |\bx_{12}|^2 u^2}{u(2u^2-\mu)(a^2 e^2+\kappa^2 u^2)}\hat{\bx}_{12} \\
    - \frac{a^2e^2 |\bx_{12}| z_{12}}{u(2u^2-\mu)(a^2e^2 + \kappa^2 u^2)}(\boldsymbol{\delta} - \hat{\bx}_{21}\hat{\bx}_{21})\cdot\bp
    \Bigg\},
\end{multline}
where $u$ and $\mu$ are given by equation \eqref{eq:uvEqn}.
Note that the second term in the right hand side of equations \eqref{eq:FR1} and \eqref{eq:FR1oblate} are the non-central parts which arise due to the anisotropy of the systems and contribute to the electrostatic torques. 
Because these force expression are valid only for large separations, they fail to account for the attractive forces between like charges that arise at short distances due to electrostatic induction.

Similarly, one can obtain a closed form expression of force using equation \eqref{eq:Feqn} and the second reflection corrections to the potential matrix (equations \eqref{eq:Phi1Ref2Rod} and \eqref{eq:Phi1Ref2Disk}) which is reliable upto $s_{21} = (|\bx_{21}| - d_{\text{min}}(\psi))/a \sim 1$. The force from the second reflection can explain the attractive interaction between like charges; however, its accuracy diminishes at the separations where the attractive region begins.  

\subsection{Electrostatic Torque}

\begin{figure*}[t] % The '*' makes it span both columns
    \centering
    \begin{subfigure}[b]{0.24\textwidth} % 1/4 of the text width minus some spacing
        \includegraphics[width=\textwidth]{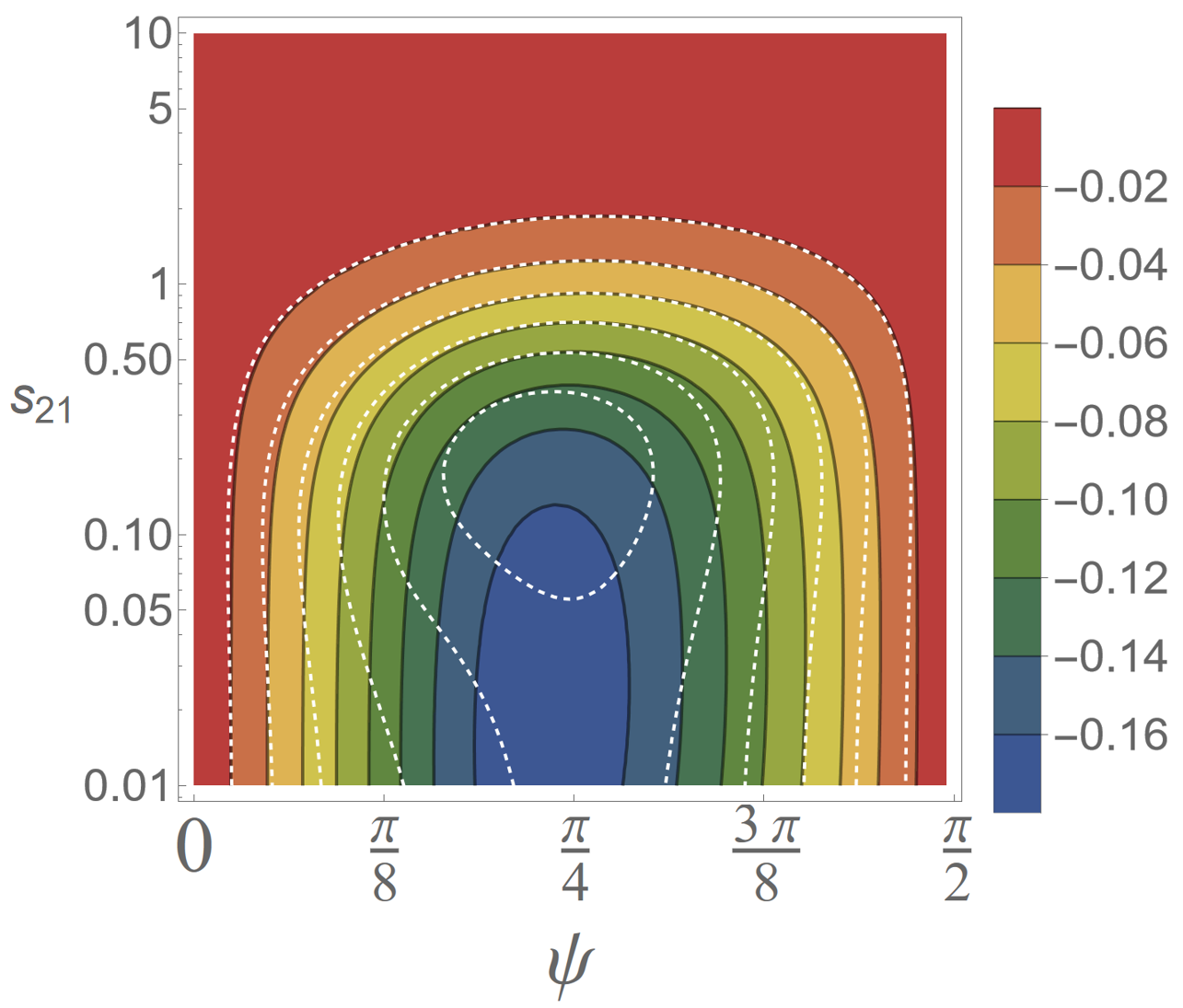}
        \caption{The prolate spheroid has charge $Q_1=q$ and the sphere has charge $Q_2=q$.}
        \label{fig:T1prolate11}
    \end{subfigure}
    \hfill
    \begin{subfigure}[b]{0.24\textwidth}
        \includegraphics[width=\textwidth]{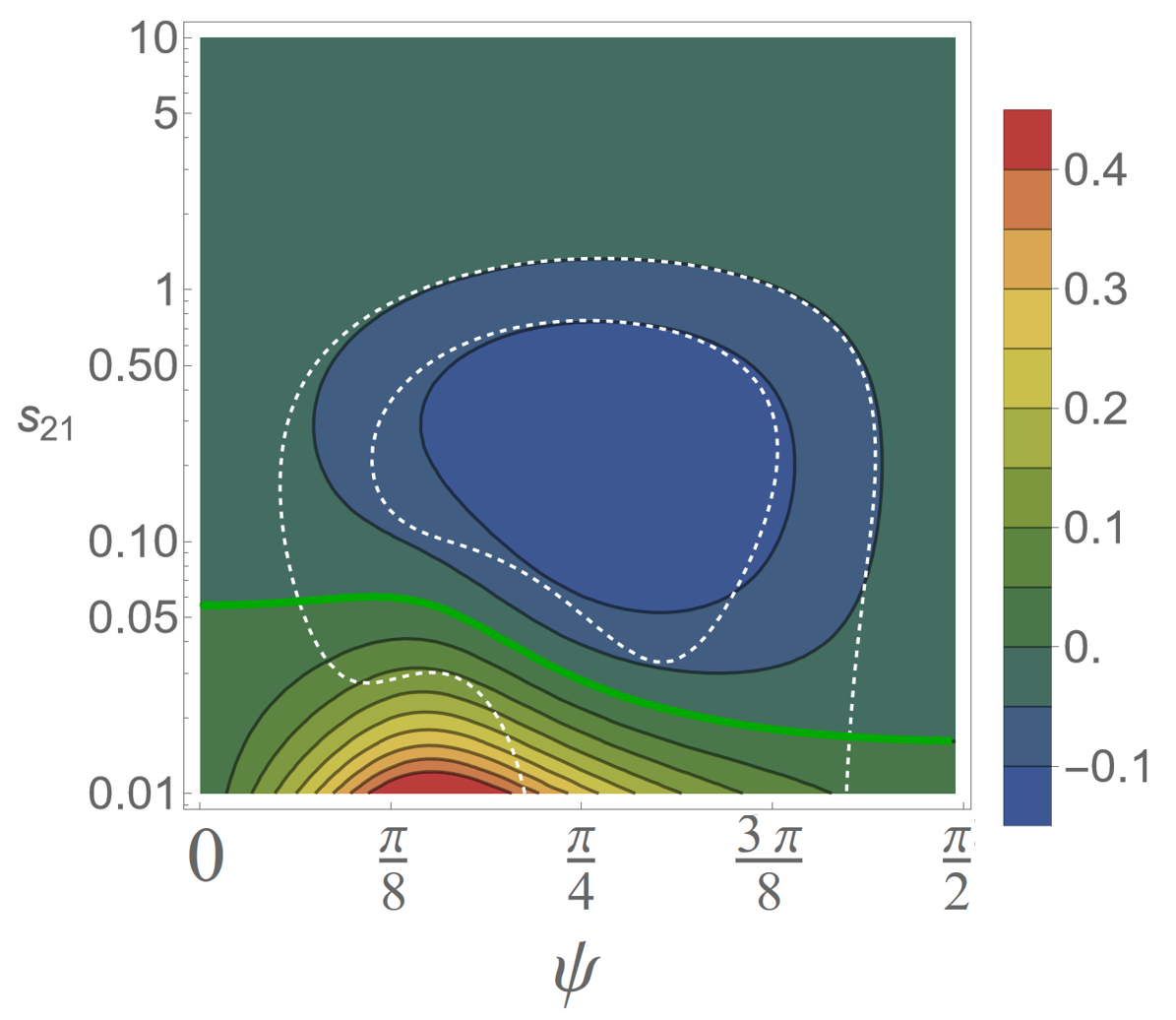}
        \caption{The prolate spheroid has charge $Q_1=q$ and the sphere has charge $Q_2=1.5q$.}
        \label{fig:T1prolate11D5}
    \end{subfigure}
    \hfill
    \begin{subfigure}[b]{0.24\textwidth}
        \includegraphics[width=\textwidth]{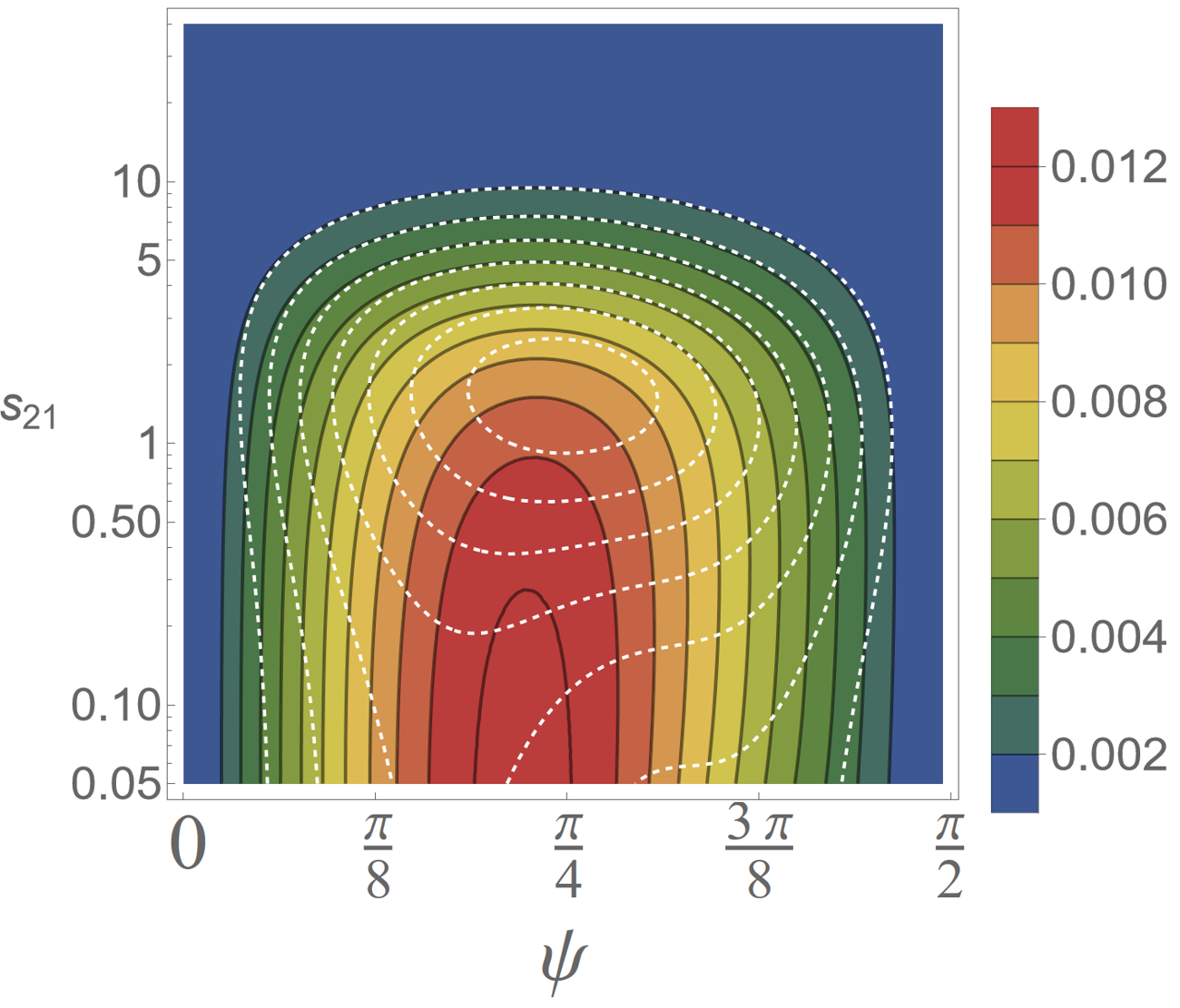}
        \caption{The oblate spheroid has charge $Q_1=q$ and the sphere has charge $Q_2=q$.}
        \label{fig:T1oblate11}
    \end{subfigure}
    \hfill
    \begin{subfigure}[b]{0.24\textwidth}
        \includegraphics[width=\textwidth]{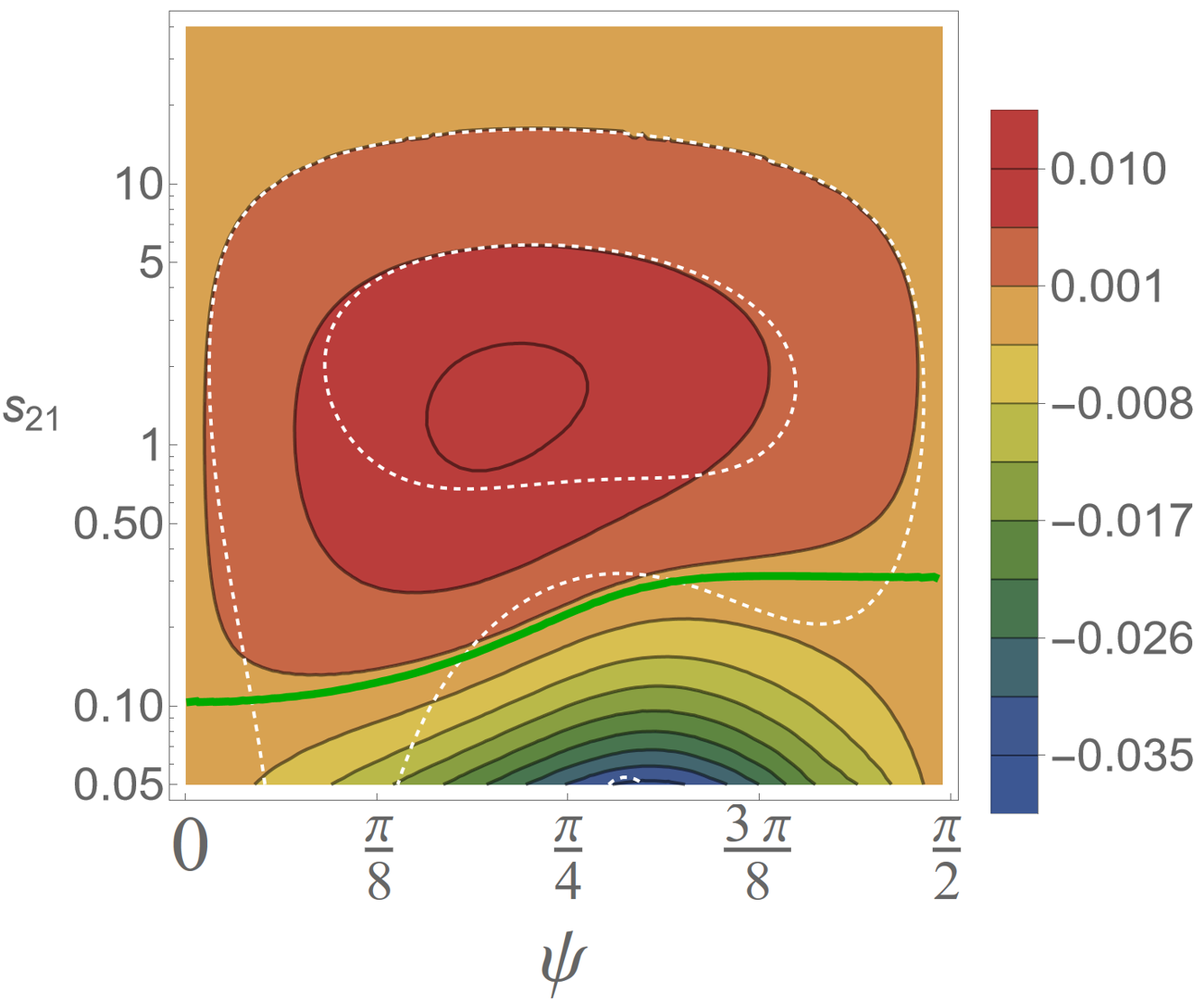}
        \caption{The oblate spheroid has charge $Q_1=q$ and the sphere has charge $Q_2=1.5q$.}
        \label{fig:T1oblate11D5}
    \end{subfigure}
    \caption{Contour plot of dimensionless torque on spheroids about their centre, $4\pi\varepsilon_0 a q^{-2}\boldsymbol{T}_1\cdot\hat{\boldsymbol{k}}$, as a function of $\psi\equiv \arccos(\hat{\bx}_{21}\cdot \bp)$ and $s_{21}=(|\bx_{21}| - d_{\text{min}}(\psi))/a$, where $\hat{\bm{k}}$ is a unit vector along $(\bp\times\hat{\bx}_{21})$. The white dotted lines are the due to the second reflections. The green curves in \ref{fig:T1prolate11D5} and \ref{fig:T1oblate11D5} separates the repulsive and the attractive regions.}
    \label{fig:Torque}
\end{figure*}

The electrostatic torque is the result of electrostatic forces on the conductors not being central. In other words, there is electrostatic energy cost in changing the orientation of the spheroid or changing the relative configuration $\psi$. Figure \ref{fig:Torque} shows the torque on the spheroid as a function of dimensionless separation $s_{21}$ and $\psi$. As the separation decreases, the torque in the unequal charge case changes direction, indicating the onset of an attractive interaction between the conductors.

\begin{figure}[H]
	\includegraphics[width=1.0\columnwidth]{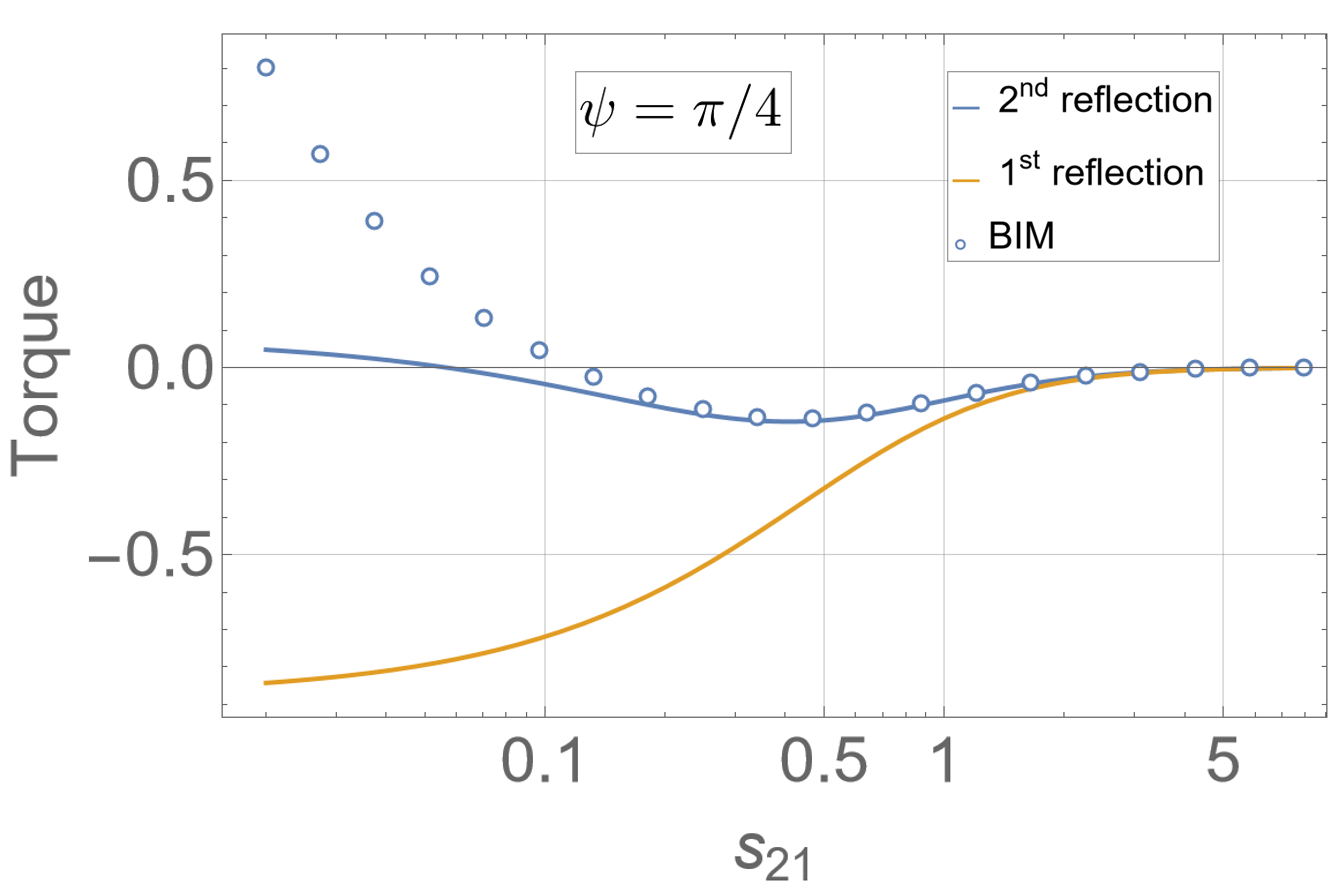}
	\caption{Dimensionless torque on the prolate spheroid $4\pi\varepsilon_0 a q^{-2}\boldsymbol{T}_1\cdot \hat{\bm{k}}$ for $Q_1=q,\, Q_2=2q$, as a function of separation $s_{21}=(|\bx_{21}| - d_{\text{min}}(\psi))/a$ for a fixed $\psi\equiv \arccos(\hat{\bx}_{21}\cdot \bp) = \pi/4$, where $\hat{\bm{k}}$ is a unit vector along $(\bp\times\hat{\bx}_{21})$. The method of reflections aligns well with the BIM in the far field. The sign change in the torque at close range indicates an attractive electrostatic force due to induction. While the first reflection fails to predict this sign change, the second reflection captures it but loses accuracy in this close range.}
	\label{fig:torqueVsD}
\end{figure}

A quantity of interest in the dilute regime is the electrostatic torque between pair of particles. The torque computed using the first reflection is accurate enough to capture the anisotropic effects in the far field. Using equation \eqref{eq:FR1}, one can obtain the torque for the prolate spheroid and sphere system, given by  
\begin{align}
    \label{eq:TR1}
    \boldsymbol{T}_1 \sim \frac{Q_1 Q_2}{8\pi\varepsilon_0 a e}\Bigg(& \frac{1 - ae/R_+}{R_+-ae-z_{12}} \nonumber\\
    -& \frac{1 + ae/R_-}{R_- + ae-z_{12}} \Bigg) \bp\times \bx_{21}.
\end{align}
Similarly, using equation \eqref{eq:FR1oblate}, one can obtain the torque for the oblate spheroid and sphere system, given by  
\begin{equation}
    \label{eq:TR1Oblate}
    \boldsymbol{T}_1 \sim \frac{Q_1 Q_2}{4\pi\varepsilon_0 u(2u^2-\mu)}\Bigg(\frac{-a^2e^2 z_{12}}{(a^2e^2 + \kappa^2 u^2)}\Bigg) \bp\times\bx_{21}.
\end{equation}
Here $R_-, R_+, z_{12}, u$ and $\mu$ are given by equations \eqref{eq:R1R2} and \eqref{eq:uvEqn}.
Note that the electrostatic forces and torques upto first reflection do not depend on the radius of the sphere\footnote{Although the accuracy of the expression requires the radius of the sphere to be much less than the separation between the particles.}. This is because the electric field of a sphere, to a leading order in the far field regime, is identical to that of a point charge placed at the center of the sphere. Now, if one has a pair of spheroids in the far field regime, the electrostatic field of a spheroid can be approximated by the field due to a point charge located at its centre. Therefore, in the far field regime, the force and torque expressions (equations \eqref{eq:FR1}, \eqref{eq:FR1oblate}, \eqref{eq:TR1} and \eqref{eq:TR1Oblate}) serve as good approximations even for a spheroid-spheroid system. The comparison between torque due to first and second reflections and BIM is shown in figure \ref{fig:torqueVsD}

Studies have shown that electrostatic interactions, when combined with hydrodynamic interactions, can result in stable configurations for a pair of spheres \cite{trombley2018stable}. An array of spheres and spheroids, as well as a dilute suspension of hydrodynamically interacting spheroids, have been found to be unstable to density perturbations \cite{crowley1971viscosity, chajwa2020waves, koch1989instability}. The potential role of electrostatics in altering the stability of such systems remains unexplored.

\begin{figure}[H]
	\includegraphics[width=1.0\columnwidth]{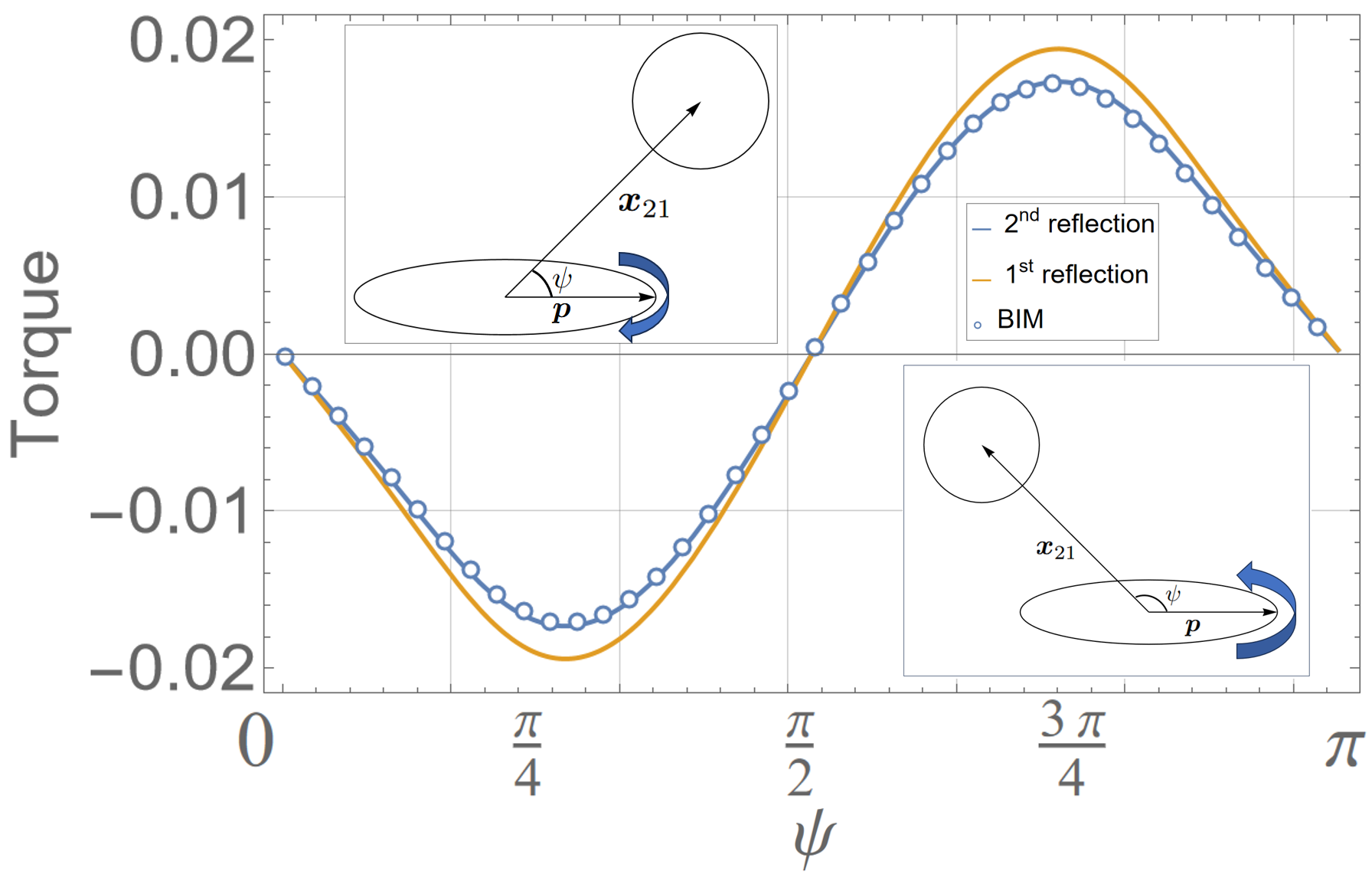}
	\caption{Dimensionless torque on the prolate spheroid $4\pi\varepsilon_0 a q^{-2}\boldsymbol{T}_1\cdot \hat{\bm{k}}$ for $Q_1=Q_2=q$, as a function of $\psi\equiv \arccos(\hat{\bx}_{21}\cdot \bp)$ for fixed $s_{21}=(|\bx_{21}| - d_{\text{min}}(\psi))/a = 2$, where $\hat{\bm{k}}$ is a unit vector along $(\bp\times\hat{\bx}_{21})$. The change in the sign of the torque shows a stable configuration of the prolate spheroid and sphere system about $\psi=\pi/2$, as indicated in the insets.}
	\label{fig:torqueStable}
\end{figure}

\begin{figure}[H]
	\includegraphics[width=1.0\columnwidth]{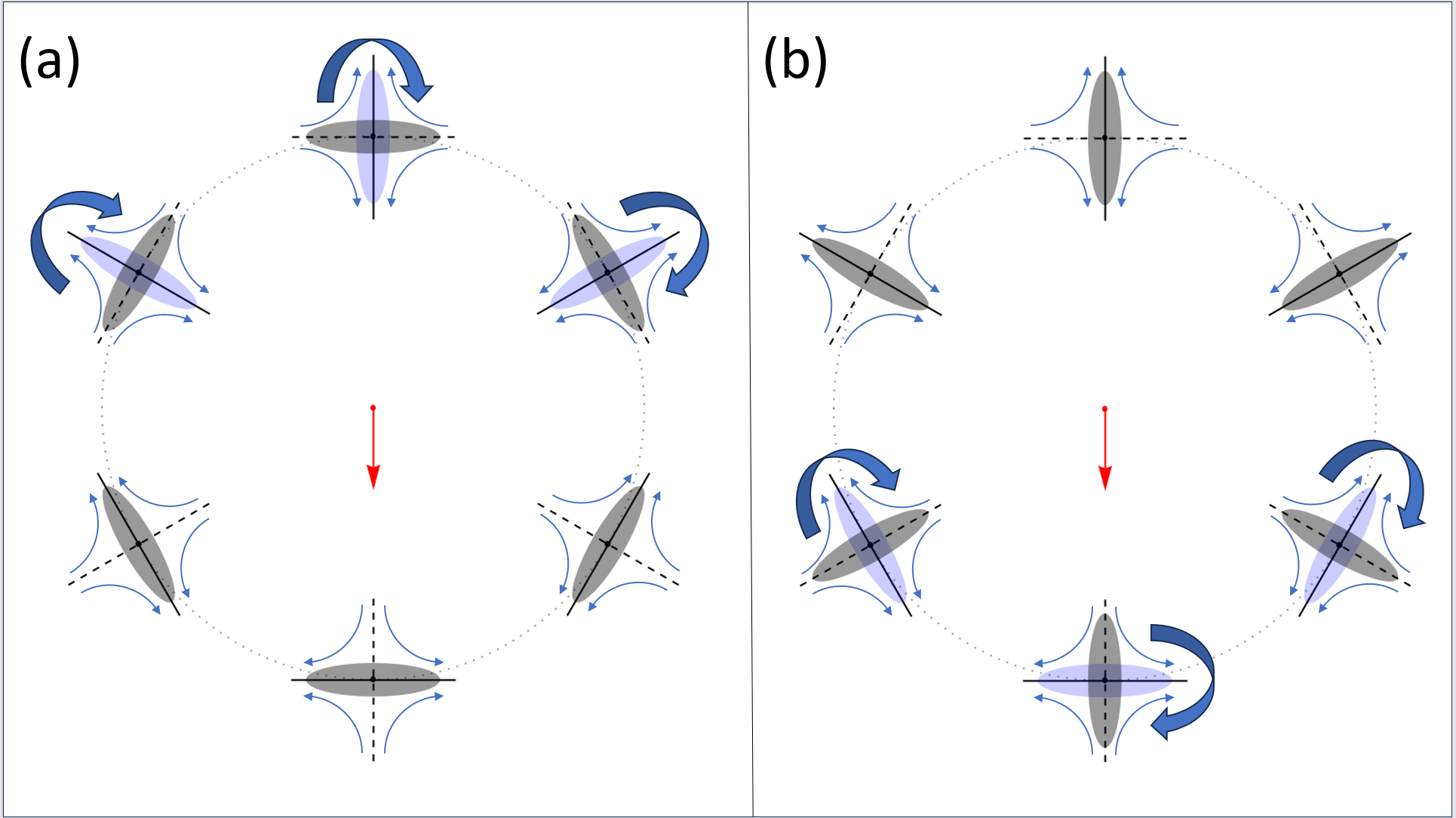}
	\caption{Schematic showing the favorable orientations of a sedimenting spheroid interacting with another sedimenting spheroid through electrostatic and hydrodynamic interactions in the far-field regime. In the case of purely hydrodynamic interactions, one spheroid disturbs the flow as a force monopole (indicated by the red arrow) and causes the other spheroid to align along the extensional axis of the locally disturbed strain field (indicated by blue arrows). When electrostatic interactions are included, the electrostatic torque can either compete with or reinforce the hydrodynamic alignment, depending on whether the spheroid is in a trailing or leading position. The black-shaded spheroids represent the favorable orientations due to electrostatic effects, while the light blue-shaded spheroids indicate those due to hydrodynamic effects.
    (a) For like-charged spheroids, the electrostatic torque competes with the hydrodynamic alignment for a trailing spheroid, as indicated by the arrows, while it reinforces the alignment for a leading spheroid. (b) For oppositely charged spheroids, the effects are reversed: the electrostatic torque competes with the hydrodynamic alignment for a leading spheroid and reinforces it for a trailing spheroid. This has implications in changing the stability of dilute suspension of charged spheroids.}
 	\label{fig:kochSchematic}
\end{figure}

In the like-charged anisotropic system, the electrostatic torque tends to align the spheroid in a broad-side orientation relative to the separation vector $\bx_{21}$, as illustrated in figure \ref{fig:torqueStable}. In contrast, for oppositely charged particles, the stable orientation changes to thin side, as evident from equations \eqref{eq:TR1} and \eqref{eq:TR1Oblate}.
% In the spheroid-sphere system, the electrostatic torque aligns the relative configuration to $\psi=\pi/2$ for a prolate spheroid and $\psi=0$ for an oblate spheroid in the case of like charges, as illustrated in figure \ref{fig:torqueStable}. 
These stable configurations contrast with the same system interacting hydrodynamically in a viscous flow \cite{koch1989instability}, where a spheroid falling above another one tends to align its thin side along their separation vector, see figure \ref{fig:kochSchematic}. 
As a result, in a dilute suspension of sedimenting charged spheroids, the hydrodynamic torque on a spheroid counteracts the electrostatic torque in some regions while reinforcing it in others. Consequently, incorporating electrostatic effects in such systems could alter the instability typically observed in purely hydrodynamic interactions \cite{koch1989instability}. 

\section{Conclusion}

We have used the method of reflections to compute the potential matrix for sphere-sphere and spheroid-sphere conductors. This allows us to determine the electrostatic forces and torques acting on these conductors in the far field regime. The formulation is general enough to be applied to arbitrary shapes as long as their singularity solutions are known, as discussed in the appendix. We also compute the electrostatic force under the lubrication approximation for nearly touching conductors in the axisymmetric configuration. To determine this close range force accurately an order one constant $\delta$ is needed which has been determined using the Boundary Integral Method (BIM). We also test the validity of the method of reflections with the BIM when the conductors are closely separated. The results show that second reflection works well until the separation is of the order of the size of the conductors. 

The anisotropy of the problem of electrostatic interaction between a spheroid and a sphere results in the electrostatic torque. This torque tends to align the spheroid-sphere system in a manner different from the alignment due to pure hydrodynamic interactions \cite{koch1989instability}, see figure \ref{fig:kochSchematic}. This naturally prompts the question: how does the instability in a dilute suspension of sedimenting spheroids change when electrostatic effects are taken into account? 
Our work offers a foundational approach for computing electrostatic forces and torques on anisotropic particle pairs, demonstrated with example cases for a spheroid-sphere system, using the potential matrix.
In the dilute regime, the simpler first-reflection expressions (equation \eqref{eq:TR1} and \eqref{eq:TR1Oblate}) can be used to account for electrostatic interactions between spheroids and study the evolution of density perturbations in a spheroid suspension.   

This work draws extensively on concepts from micro-hydrodynamics but deliberately excludes its effects to avoid additional complexity. However, in natural settings, micro-hydrodynamics and electrostatic effects often act together. Understanding the role of electrostatic forces in clustering within clouds, for instance, sheds light on the formation and dynamics of ice crystals and droplets.
While hydrodynamic-driven clustering through turbulence has been extensively explored \cite{bec2014gravity, meibohm2021paths}, the role of electrostatic interactions remains under-examined. Such insights can further our understanding of processes such as rain initiation, hail formation, and the structural evolution of clouds under varying atmospheric charge distributions. Beyond atmospheric science, applications extend to areas like the control of particulate matter in industrial filtration \cite{wang2001electrostatic}, the alignment of particles in electric fields in colloidal chemistry \cite{edwards2014controlling}, and the behavior of charged proteins in biophysics \cite{sun2022electrostatics}, where electrostatic torques influence assembly and organization.

\begin{acknowledgements}
HJ acknowledges support of the Department of Atomic Energy, Government of India, under project no. RTI4001.
\end{acknowledgements}
\nocite{*}

\appendix

\section{Singularity solutions for spheroids in electrostatics}

The singularity solutions for the boundary value problems of the Laplace equation involves representing the solution in terms of the Greens function $\mathcal{G}$ of the Laplace equation and its higher derivatives located outside the domain of interest. The Greens function of the free space Laplace equation in 3D satisfies 
\begin{equation}
    \label{eq:GfnEqn}
    \nabla^2\mathcal{G}(\bx) = -\delta(\bx)
\end{equation}
and is given by
\begin{equation}
    \label{eq:Gfn}
    \mathcal{G}(\bx) = \frac{1}{4\pi |\bx|}.
\end{equation}

\subsection{Charged prolate spheroid}

Any point $\bx$ on a prolate spheroid $S_p$ with semi-major axis $a$ and aspect ratio $\kappa (>1)$, oriented along the unit vector $\bp$ and centered at origin is given by
\begin{equation}
    \label{eq:prolateEqn}
    \bx\cdot\left[\frac{1}{a^2}\bp\bp + \frac{1}{a^2 \kappa^{-2}}(\boldsymbol{\delta}-\bp\bp)\right]\cdot\bx = 1,\quad \bx \in S_p.
\end{equation}
The boundary value problem to be solved for the potential field outside $S_p$ is
\begin{subequations}
\label{eq:ChargeEqnRod}
\begin{gather}
    \nabla^2 \phi(\bx) = 0, \\
    \phi(\bx) = \phi_0, \quad \bx \in S_p\\
    \phi(\bx) \to 0 \quad \text{as}\quad |\bx| \to \infty.
\end{gather}    
\end{subequations}
The solution can be represented in terms of a uniform charge distribution located along the symmetry axis of $S_p$ as \cite{alawneh1977singularity}
\begin{equation}
    \label{eq:solRodQ}
    \phi(\bx) = \phi_0\left\{\frac{2\pi}{\arctanh e}\int_{-ae}^{ae} \mathcal{G}(\bx-\xi \bp) \, d\xi\right\},    
\end{equation}
where $e=\sqrt{1-b^2/a^2}$ is the eccentricity. The total charge $Q$ on the surface of $S_p$ is given by
\begin{align}
    \label{eq:QCVeqnRod}
    Q = -\varepsilon_0\oint_{S_p} \boldsymbol{\hat n}\cdot \boldsymbol{\nabla}\phi\, dS =& -\varepsilon_0\int_{V_p} \nabla^2\phi\, d\tau \nonumber \\
    =& \left[ \frac{4\pi a \varepsilon_0 e}{\arctanh e} \right] \phi_0,
\end{align}
where $\varepsilon_0$ is the permittivity of the free space and $\boldsymbol{\hat n}$ is the outward normal vector to $S_p$ and $V_p$ is the volume inside $S_p$. Therefore, the capacitance $C \equiv Q/\phi_0$ of the perfectly conducting prolate spheroid $S_p$ is given by \cite{landau2013electrodynamics, zangwill2013modern, lekner2021electrostatics, stratton2007electromagnetic}
\begin{equation}
    \label{eq:CeqnRod}
    C = \frac{4\pi a \varepsilon_0 e}{\arctanh e}.
\end{equation}
Note that as $\lim_{e\to 0}C = 4\pi a\varepsilon_0$, which is the capacitance of a sphere of radius $a$.

\subsection{Charged oblate spheroid}

The singularity solution of an oblate spheroid can be derived from that of a prolate spheroid using the eccentricity transformation \cite{shatz2004singularity}
\begin{equation}
    \label{eq:eTransfrm}
    e \to \dfrac{i e}{\sqrt{1-e^2}}.
\end{equation}
Therefore, the potential field due to an isolated oblate spheroid described by equation \eqref{eq:prolateEqn} with $\kappa < 1$ is given by:
\begin{equation}
    \label{eq:solDiskQ}
    \phi(\bx) = \phi_0\left\{ \frac{2\pi}{\arcsin e} \int_{-ae/\kappa}^{ae/\kappa} \mathcal{G}(\bx-i \xi \bp) \, d\xi \right\}.
\end{equation}
Note that for cartesian coordinates aligned such that the unit vector $\bp$ is along the z-axis, $\mathcal{G}(\bx-i\xi \bp)$ gives rise to a term $\dfrac{1}{\sqrt{x^2+y^2+(z-i\xi)^2}}$ which is singular on the disk of radius $\xi$ in the $x-y$ plane ($z=0$), which corresponds to the singularity distribution for an oblate spheroid \cite{kim2013microhydrodynamics}.

Correspondingly, the capacitance of an isolated oblate spheroid is given by \cite{landau2013electrodynamics, zangwill2013modern, lekner2021electrostatics, stratton2007electromagnetic}
\begin{equation}
    \label{eq:CeqnDisk}
    C = \frac{4\pi a\varepsilon_0 e}{\kappa\arcsin e}.
\end{equation}

\subsection{Grounded prolate spheroid in presence of a uniform electric field}

The potential field in this case can be divided into two parts as $\phi = \phi^d + \phi^\infty$. Here $\phi^d$ is the disturbance potential produced by the grounded prolate spheroid so as to maintain zero potential on its surface and $\phi^\infty = -\boldsymbol{E}^\infty\cdot \bx$, with $\boldsymbol{E}^\infty$ being the ambient uniform electric field. The boundary value problem to be solved for $\phi^d(\bx)$ outside $S_p$ in this case is
\begin{subequations}
\label{eq:diffEqnrod}
\begin{gather}
    \nabla^2\phi^d(\bx) = 0,\\
    \phi^d(\bx) = \boldsymbol{E}^\infty\cdot \bx, \quad \bx \in S_p,\\
    \phi^d(\bx)\to 0 \quad \text{as} \quad |\bx| \to \infty.
\end{gather}
\end{subequations}
The solution can be represented as \cite{alawneh1977singularity}
\begin{multline}    
    \label{eq:solRodD}
    \phi^d(\bx) = \boldsymbol{E}^\infty\cdot \Bigg\{\frac{6\pi X^C_p}{e^3}\bp \int_{-ae}^{ae} \xi\, \mathcal{G}(\bx - \xi \bp)\, d\xi
    \\- \frac{3\pi Y^C_p}{e^3}(\boldsymbol{\delta}-\bp\bp)\cdot\boldsymbol{\nabla}\int_{-ae}^{ae} (a^2e^2-\xi^2) \mathcal{G}(\bx-\xi\bp)\, d\xi \Bigg\},
\end{multline}
where 
\begin{subequations}
    \label{eq:XcYc}
\begin{gather}
    X^C_p \equiv \frac{e^3}{3}(\arctanh e-e)^{-1}, \\ Y^C_p \equiv \frac{2e^3}{3}\left( \frac{e}{1-e^2}-\arctanh e \right)^{-1}. 
\end{gather}
\end{subequations}
The first integral term in equation \eqref{eq:solRodD} represents a linear charge distribution along the symmetry axis whereas the second integral term represents the parabolic distribution of dipole moments pointing perpendicular to the symmetry axis. Note that the charge distribution in the first integral term has non-zero dipole moment but zero net charge. The induced dipole moment $\boldsymbol{d}$ is given by
\begin{equation}
    \label{eq:dipoleRodTmp}
    \boldsymbol{d} = -\varepsilon_0\oint_{S_p} \bx \, \boldsymbol{\hat n}\cdot \boldsymbol{\nabla}\phi\, dS =  -\varepsilon_0\int_{V_p} [\boldsymbol{\nabla}\phi + \bx\, \nabla^2\phi]\, d\tau
\end{equation}
The volume integral of the gradient term doesn't contribute since $\phi = 0$ on $S_p$ and
\begin{equation}
    \label{eq:gradTerm}
    \int_{V_p} \boldsymbol{\nabla}\phi\, d\tau = \oint_{S_p} \phi\, \boldsymbol{\hat n}\, dS = 0.
\end{equation}
Therefore, the dipole moment is given by
\begin{align}
    \label{eq:dipoleRod}
    \boldsymbol{d} =&  -\varepsilon_0\int_{V_p} \bx\, \nabla^2\phi^d\, d\tau \nonumber \\
    =& \quad 4\pi a^3 \varepsilon_0[X^C_p\bp \bp + Y^C_p(\boldsymbol{\delta} - \bp\bp)]\cdot \boldsymbol{E}^\infty.
\end{align}
Note that $\lim_{e\to0}X^C_p = \lim_{e\to0} Y^C_p = 1$ resulting in $\lim_{e\to 0} \boldsymbol{d}=4\pi a^3\varepsilon_0\boldsymbol{E}^\infty$ and we get the dipole moment of a sphere of radius $a$. We can rewrite equation \eqref{eq:solRodD} in terms of the dipole moment $\boldsymbol{d}$ as
\begin{multline}
    \label{eq:solRodD2}
    \phi^d(\bx) = \frac{3}{2a^3e^3\varepsilon_0} \boldsymbol{d}\cdot\bp\int_{-ae}^{ae} \xi\, \mathcal{G}(\bx-\xi\bp)\, d\xi \\
    - \frac{3}{4a^3e^3\varepsilon_0} \boldsymbol{d}\cdot(\boldsymbol{\delta}-\bp\bp)\cdot \boldsymbol{\nabla}_{\bx} \int_{-ae}^{ae} (a^2e^2-\xi^2)\, \mathcal{G}(\bx-\xi\bp)\, d\xi.
\end{multline}

\subsection{Grounded oblate spheroid in presence of a uniform electric field}

We again use the eccentricity transformation \eqref{eq:eTransfrm} to obtain the dipole moment $\boldsymbol{d}$ and disturbance potential field $\phi^d(\bx)$ due to a grounded oblate spheroid in presence of a uniform background electric field $\boldsymbol{E}^\infty$. The dipole moment is given by
\begin{equation}
    \label{eq:dipoleDisk}
    \boldsymbol{d} = 4\pi a^3 \varepsilon_0[X^C_o\bp \bp + Y^C_o(\boldsymbol{\delta} - \bp\bp)]\cdot \boldsymbol{E}^\infty,
\end{equation}
where 
\begin{subequations}
\label{eq:XcYcDisk}
\begin{gather}
    X^C_o \equiv \frac{e^3}{3}[e(1-e^2) - (1-e^2)^{3/2}\arcsin e]^{-1},\\ 
    Y^C_o \equiv \frac{2e^3}{3}\left[ e(1-e^2)^2 - (1-e^2)^{3/2}\arcsin e \right]^{-1}. 
\end{gather}
\end{subequations}
The disturbance potential field is given by:
\begin{multline}
    \label{eq:solDiskD2}
    \phi^d(\bx) = \frac{3\kappa^3}{2a^3e^3\varepsilon_0} \bigg\{\boldsymbol{d}\cdot\bp\int_{-ae/\kappa}^{ae/\kappa} -i    \xi\, \mathcal{G}(\bx-i\xi\bp)\, d\xi \\- \frac{1}{2} \boldsymbol{d}\cdot(\boldsymbol{\delta}-\bp\bp)\cdot \boldsymbol{\nabla}_{\bx} \int_{-ae/\kappa}^{ae/\kappa} \left(\frac{a^2e^2}{\kappa^2}-\xi^2\right)\, \mathcal{G}(\bx-i\xi\bp)\, d\xi \bigg\}.
\end{multline}

\section{Faxén Laws for arbitrary shaped conductors in electrostatics}
\label{app:faxenLaws}

Faxén laws for electrostatics can be derived analogously to those in microhydrodynamics \cite{kim2013microhydrodynamics}, using the reciprocal theorem. The electrostatic counterpart for spheres is detailed in \cite{bonnecaze1990method}, and we extend this framework to arbitrarily shaped conductors. 
Let $\phi_1$ and $\phi_2$ be two fields in the same domain $D$. The reciprocal theorem states that
\begin{align}
    \label{eq:reciprocalThm}
    \int_D \phi_2 \nabla^2 \phi_1 \, d\tau - \int_D \phi_1 \nabla^2 \phi_2 \, d\tau =& \oint_{\partial D} \phi_2 \boldsymbol{\nabla}\phi_1\cdot\boldsymbol{n}\, dS \nonumber \\
    -& \oint_{\partial D} \phi_1 \boldsymbol{\nabla}\phi_2\cdot\boldsymbol{n}\, dS, 
\end{align}
where $\boldsymbol{n}$ is the normal vector to the boundary of the domain $D$, denoted by $\partial D$, pointing away from $D$. In the context of a conductor placed in a potential field, $D$ is the region in $\mathbb{R}^3$ bounded by the surface of the conductor and a large sphere ``at infinity''.

\subsection{Faxén law for total charge and potential on a conductor}
\label{sec:faxen1Law}
We follow the approach of \cite{kim2013microhydrodynamics} to relate the total charge $Q$ on the surface of a conductor to its surface potential $V$ in presence of an arbitrary background potential field $\boldsymbol{\phi^\infty}(\bx)$, such that $\phi^\infty(\bx) \sim \mathcal{O}(1/|\bx|)$ as $\bx$ goes to infinity. 
Let us denote the surface of an arbitrary shaped conductor by $S_p$. 
Note that total charge on $S_p$ due to a potential $\phi(\bx)$ outside it, is given by
\begin{equation}
    \label{eq:chargePhi}
    Q = -\varepsilon_0 \oint_{S_p} \boldsymbol{\nabla}\phi\cdot\boldsymbol{\hat n}\, dS,
\end{equation}
where $\boldsymbol{\hat n}$ is the outward pointing normal vector to $S_p$ and $\varepsilon_0$ is the permittivity of free space.

We use the reciprocal theorem with the details of the two fields as follows: 
\begin{enumerate}
    \item Take $\phi_1$ to be the potential field satisfying the Laplace equation outside the isolated conductor with $\phi_1=\phi_{10}$ on $S_p$, where $\phi_{10}$ is some constant and $\phi_1$ goes to zero at infinity. 
    This is a case of an isolated conductor with some charge $Q_1$ on its surface given by $Q_1 = C \phi_{10}$, where $C$ is the capacitance of the conductor.
    \item  Take $\phi_2$ to be the potential field given by the solution of $\nabla^2 \phi_2(\bx) = -Q' \varepsilon_0^{-1}\, \delta(\bx - \by)$, where $\by\in D$, with $\phi_2 = V$ on $S_p$. 
    Here the ambient potential field $\phi^\infty_2(\bx)$ is given by a point charge located at $\by$ and the conductor produces a disturbance field in order to satisfy the boundary condition on its surface. 
    Let $Q_2$ be the charge on the conductor, which is to be determined using the reciprocal theorem.
\end{enumerate}

Using equations \eqref{eq:reciprocalThm} and \eqref{eq:chargePhi}, we have\footnote{We have used the fact that the fields involved in the problems decay fast enough far from the conductor so as to have zero contribution from the surface ``at infinity''.},     
\begin{gather}
    \label{eq:F1derv}
    Q' \phi_1(\by) = Q_1 V - Q_2 \phi_{10} \nonumber\\
    \implies Q_2 \phi_{10} = C \phi_{10} V - Q' \phi_1(\by).
\end{gather}
Now, $\phi_1(\by)$ can be represented in terms of singularity solution as
\begin{equation}
    \label{eq:singPotential}
    \phi_1(\by) = \phi_{10}\mathcal{F}_V\{\mathcal{G}(\by-\boldsymbol\xi)\} = \phi_{10}\mathcal{F}_V\{\mathcal{G}(\boldsymbol\xi-\by)\}.
\end{equation}
Here $\mathcal{F}_V$ is the corresponding linear functional and $\boldsymbol{\xi}$ represents the region inside the conductor over which the singularities are distributed.
Using equations \eqref{eq:F1derv} and \eqref{eq:singPotential}, we have
\begin{equation}
    \label{eq:F1LawEg}
    Q_2 = CV - Q' \mathcal{F}_V\{\mathcal{G}(\boldsymbol{\xi}- \by) \} = CV - \mathcal{F}_V\{\phi^\infty_2(\boldsymbol{\xi}) \}.
\end{equation}
Here we have used the fact that $Q'\mathcal{G}(\boldsymbol{\xi}-\by) = \phi_2^\infty(\boldsymbol{\xi})$. However, all ambient fields $\phi^{\infty}(\bx)$ that decay at infinity and satisfy the Laplace equation can be constructed using appropriate set of point charges. Therefore, equation \eqref{eq:F1LawEg} applies to a general ambient field $\phi^{\infty}(\bx)$. Thus, the relation between charge $Q$ on a conductor and the potential $V$ on its surface in presence of a background potential field $\phi^\infty(\bx)$ is given by
\begin{equation}
    \label{eq:F1law}
    Q = CV - \mathcal{F}_V\{\phi^\infty(\boldsymbol{\xi})\}.
\end{equation}

This result can be directly applied to the bodies with known singularity solution of the form given in equation \eqref{eq:singPotential}. In particular, for a prolate spheroid with semi-major axis $a$, eccentricity $e$ and orientation vector $\bp$ we have the singularity representation given by equation \eqref{eq:solRodQ} and capacitance by equation \eqref{eq:CeqnRod}. Therefore, the charge $Q$ on the prolate spheroid in presence of a background potential field $\phi^\infty$ is given by
\begin{equation}
    \label{eq:QVrod}
    Q = \frac{4\pi a \varepsilon_0 e}{\arctanh e}\left\{V - \frac{1}{2a e}\int_{-ae}^{ae} \phi^\infty(\bx_c+\xi\bp)\, d\xi \right\},
\end{equation}
where $c=ae$ and $\bx_c$ denotes the centre of the prolate spheroid.

Similarly, the charge relation for an oblate spheroid with semi-major axis $a$ and orientation vector $\bp$ in presence of a background potential field $\phi^\infty$ is given by 
\begin{equation}
    \label{eq:QVDisk}
    Q = \frac{4\pi a \varepsilon_0 e}{\kappa\arcsin e}\left\{V - \frac{\kappa}{2a e}\int_{-ae/\kappa}^{ae/\kappa} \phi^\infty(\bx_c+i\xi\bp)\, d\xi \right\},
\end{equation}

\subsection{Faxén law for induced dipole moment on a conductor}
\label{sec:faxen2Law}
To relate the induced dipole moment $\boldsymbol{d}$ on a conductor due in presence of an ambient potential field $\phi^\infty(\bx)$, we again use the reciprocal theorem with the details of the two fields as follows:
\begin{enumerate}
    \item Take $\phi_1$ to be the potential field satisfying the Laplace equation outside the isolated conductor with $\phi_1=\boldsymbol{E}_{10}^\infty\cdot\bx$ on $S_p$, where $\boldsymbol{E}_{10}^\infty$ is a constant electric field and $\phi_1$ goes to zero at infinity. This is a case of the disturbance potential produced by a grounded isolated conductor placed in a uniform ambient field $\boldsymbol{E}^{\infty}_{10}$. 
    \item  Take $\phi_2$ to be the potential field given by the solution of $\nabla^2 \phi_2(\bx) = -Q' \varepsilon_0^{-1}\, \delta(\bx - \by)$, where $\by\in D$, with $\phi_2 = 0$ on $S_p$. 
    The goal is to determine the induced dipole moment $\boldsymbol{d}_2$ in this case.
\end{enumerate}
Applying the reciprocal theorem in these two fields gives
\begin{equation}
    \label{eq:F2Derv}
    Q'\phi_1(\by) = \varepsilon_0\boldsymbol{E}^\infty_{10}\cdot\oint_{S_p} \bx \boldsymbol{\nabla}\phi_2\cdot \boldsymbol{\hat n}\, dS = -\boldsymbol{E}^\infty_{10}\cdot\boldsymbol{d}_2,
\end{equation}
where we have used the fact that the surface charge density on the conductor is given by $\sigma_2 = -\varepsilon_0 \boldsymbol{\nabla}\phi_2.\boldsymbol{\hat n}$ and dipole moment $\boldsymbol{d}_2$ is simply the first moment of this charge density on the conductor.
Now, $\phi_1(\by)$ can be represented in terms of singularity solution as
\begin{equation}
    \label{eq:sing2Potential}
    \phi_1(\by) = \boldsymbol{E}_{10}^\infty\cdot\boldsymbol{\mathcal{F}}_E\{\mathcal{G}(\by-\boldsymbol\xi)\} = \boldsymbol{E}_{10}^\infty\cdot\boldsymbol{\mathcal{F}}_E\{\mathcal{G}(\boldsymbol\xi-\by)\}.
\end{equation}
Here $\boldsymbol{\mathcal{F}}_E$ is the corresponding linear functional and $\boldsymbol{\xi}$ represents the region inside the conductor over which the singularities are distributed. Using this in equation \eqref{eq:F2Derv} and factoring out $\boldsymbol{E}_{10}^\infty$, we have
\begin{equation}
    \label{eq:F2lawEg}
    \boldsymbol{d}_2 = -\boldsymbol{\mathcal{F}}_E\{Q'\mathcal{G}(\boldsymbol{\xi}-\by)\} = -\boldsymbol{\mathcal{F}}_E\{\phi_2^\infty(\boldsymbol{\xi})\},
\end{equation}
where $\phi_2^\infty$ is the ambient potential field in the second case. Again, for the general ambient field $\phi^\infty(\bx)$ constructed using appropriate set of point charges, the dipole moment $\boldsymbol{d}$ on the conductor is simply given by
\begin{equation}
    \label{eq:F2law}
    \boldsymbol{d} = -\boldsymbol{\mathcal{F}}_E\{\phi^\infty(\boldsymbol{\xi})\}.
\end{equation}

This result can be directly applied to the bodies with known singularity solution of the form given
in equation \eqref{eq:sing2Potential}. In particular, for a prolate spheroid with semi-major axis $a$, eccentricity $e$ and
orientation vector $\bp$ we have the singularity representation given by equation \eqref{eq:solRodD}. Therefore, the induced dipole moment on the prolate spheroid in presence of a background potential
field $\phi^\infty$ is given by
\begin{multline}
    \label{eq:dRod}
    \boldsymbol{d} = -4\pi a^3\varepsilon_0\Bigg[ \frac{3}{2a^3e^3} X^C_p \bp \int_{-ae}^{ae} \xi\, \phi^\infty(\bx_c+\xi \bp)\, d\xi 
     \\+\frac{3}{4a^3e^3} Y^C_p (\boldsymbol{\delta}-\bp\bp)\cdot \boldsymbol{\nabla}_{\bx_c} \int_{-ae}^{ae} (a^2e^2-\xi^2)\, \phi^\infty(\bx_c+\xi \bp)\, d\xi  \Bigg].
\end{multline}

Similarly, the dipole moment of an oblate spheroid with semi-major axis $a$ and orientation vector $\bp$ in presence of a background potential field $\phi^\infty$ is given by
\begin{multline}
    \label{eq:dDisk}
    \boldsymbol{d} = -4\pi a^3\varepsilon_0\frac{3\kappa^3}{2a^3e^3}\Bigg[  X^C_o \bp \int_{-ae/\kappa}^{ae/\kappa} i\xi\, \phi^\infty(\bx_c+i\xi \bp)\, d\xi 
     \\+ \frac{Y^C_o}{2} (\boldsymbol{\delta}-\bp\bp)\cdot \boldsymbol{\nabla}_{\bx_c} \int_{-ae/\kappa}^{ae/\kappa} \left(\frac{a^2e^2}{\kappa^2}-\xi^2\right)\, \phi^\infty(\bx_c+i\xi \bp)\, d\xi  \Bigg].
\end{multline}

\section{Electrostatic interactions using the Method of reflections}
\label{app:RefMethod}

The exact way to incorporate electrostatic interaction between conductors would require obtaining a harmonic potential field which satisfies the constant potential boundary conditions on the surface of each conductor. This problem is barely tractable for two spheres, and hence we need to resort to some approximate methods such as the method of reflections for more complex shapes like spheroids.

Method of reflections is an iterative scheme widely used in micro-hydrodynamics to calculate hydrodynamic interactions between widely separated bodies \cite{kim2013microhydrodynamics}. This method produces a perturbation series in terms of the order $a/R$ where $a$ is the typical size of the objects and $R$ is their typical separation. The method is described in \cite{kim2013microhydrodynamics} and is outlined for electrostatic interaction between two conductors as follows. 

In the zeroth-order approximation, the solution for two conductors (denoted $S_1$ and $S_2$) that are far apart is obtained by simply adding the potential fields of each isolated conductor, meaning the electrostatic interactions between them are ignored. Let $\phi_1$ and $\phi_2$ be two potential fields such that 
\begin{subequations}
    \label{eq:reflectionDemo1}
\begin{gather}
    \phi_1(\bx) = V_1 \quad \bx \in S_1, \\
    \phi_2(\bx) = V_2 \quad \bx \in S_2.
\end{gather}
\end{subequations}
However, $\phi=\phi_1+\phi_2$ doesn't satisfy the boundary conditions on either of the surfaces. 
Infact, the error in the boundary condition on $S_\alpha$ is $\phi_{3-\alpha}(\bx)$ which is of the order of $a/R$.
The fields $\phi_1(\bx)$ and $\phi_2(\bx)$ are called the first \textit{incident fields} on the conductors $S_2$ and $S_1$, respectively.
Now, $S_1$ produces a disturbance field $\phi_{21}$ and $S_2$ produces a disturbance field $\phi_{12}$ such that
\begin{subequations}
        \label{eq:reflectionDemo2}
\begin{gather}
    \phi_{21}(\bx) = -\phi_2(\bx) \quad \bx \in S_1, \\
    \phi_{12}(\bx) = -\phi_1(\bx) \quad \bx \in S_2.
\end{gather}
\end{subequations}
These disturbance fields are called the  \textit{reflected fields} which accounts for the correction in the boundary conditions. Now, $\phi=\phi_1+\phi_2+\phi_{21}+\phi_{12}$ is a better approximation to the complete solution because the error in the boundary conditions is now $\mathcal{O}(\phi_{12}) \sim \mathcal{O}(\phi_{21})$ which takes contributions from higher multipole moments and decays faster than $a/R$. 
This procedure can be iterated with the reflected fields from one conductor being incident on the other conductor and producing subsequent reflected fields. We shall apply this method upto second reflections in case of interacting spheroids.

\subsection{Far field interaction of two prolate spheroids}

Consider two prolate spheroids $S_1$ and $S_2$ with semi-major axes $a_1$ and $a_2$, eccentricities $e_1$ and $e_2$, position vectors $\bx_1$ and $\bx_2$ and orientations $\bp_1$ and $\bp_2$, respectively. Faxén laws (see equation \eqref{eq:QVrod}) can be used to relate the potentials $V_1$ and $V_2$ on the surfaces of the spheroids to their total charges $Q_1$ and $Q_2$, respectively. The ambient field around the first spheroid is generated by the second spheroid and can be expressed perturbatively using the method of reflections. The same approach applies to the second spheroid, where its ambient field is influenced by the first spheroid. Using equation \eqref{eq:QVrod}, we have for the first spheroid
\begin{subequations}
        \label{eq:VQrflctnRod}
\begin{gather}
    V_{1} = Q_1\frac{\arctanh e_1}{4\pi a_1\varepsilon_0 e_1} + \frac{1}{2a_1e_1}\int_{-a_1e_1}^{a_1e_1} \phi_{2}^\infty(\bx_1+\xi_1\bp_1)\, d\xi_1,\\
    V_{2} = Q_2\frac{\arctanh e_2}{4\pi a_2\varepsilon_0 e_2} + \frac{1}{2a_2e_2}\int_{-a_2e_2}^{a_2e_2} \phi_{1}^\infty(\bx_2+\xi_1\bp_2)\, d\xi_1,
\end{gather}
\end{subequations}
Using the method of reflections, we have
\begin{subequations}
        \label{eq:phi2Inf}
\begin{gather}
    \phi_{1}^\infty(\by) = \phi_{1}(\by) + \phi_{21}(\by) + \phi_{121}(\by) ...\\
    \phi_{2}^\infty(\by) = \phi_{2}(\by) + \phi_{12}(\by) + \phi_{212}(\by)...
\end{gather}
\end{subequations}
Here $\phi_{1}(\by)$ and $\phi_{2}(\by)$ are the zeroth-order disturbance fields, $\phi_{21}(\by)$ and $\phi_{12}(\by)$ are the first reflection fields and $\phi_{121}(\by)$ and $\phi_{212}(\by)$ are the second reflection fields produced by $S_1$ and $S_2$, respectively.

The zeroth order solution to the problem is
\begin{equation}
    \label{eq:V0Rods}
    V_1^{(0)} = Q_1\frac{\arctanh e_1}{4\pi a_1\varepsilon_0 e_1}, \quad V_2^{(0)} = Q_2\frac{\arctanh e_2}{4\pi a_2\varepsilon_0 e_2}.
\end{equation}
 Since $\phi_1$ and $\phi_2$ are the potentials due to isolated spheroids $S_1$ and $S_2$, they are given by equations \eqref{eq:solRodQ} and \eqref{eq:QCVeqnRod} as
\begin{subequations}
        \label{eq:phi2Eqn}
\begin{gather}
    \phi_1(\by) = \frac{Q_1}{2\varepsilon_0a_1e_1}\int_{-a_1e_1}^{a_1e_1} \mathcal{G}(\by-\bx_1-\xi_1\bp_1)\, d\xi_1,\\
    \phi_2(\by) = \frac{Q_2}{2\varepsilon_0a_2e_2}\int_{-a_2e_2}^{a_2e_2} \mathcal{G}(\by-\bx_2-\xi_1\bp_2)\, d\xi_1.
\end{gather}
\end{subequations}
The first order correction comes through the first reflection as 
\begin{subequations}
 \label{eq:V1Rods}
   \begin{gather}
    V_1^{(1)} = \frac{1}{2a_1e_1}\int_{-a_1e_1}^{a_1e_1} \phi_{2}(\bx_1+\xi_1\bp_1)\, d\xi_1, \\ 
    V_2^{(1)} = \frac{1}{2a_2e_2}\int_{-a_2e_2}^{a_2e_2} \phi_{1}(\bx_2+\xi_1\bp_2)\, d\xi_1,
\end{gather}
\end{subequations}
with the first reflection fields $\phi_{21}$ and $\phi_{12}$ represented to the leading order in $a/R$ by the dipole moments $\boldsymbol{d}_1^{(1)}$ and $\boldsymbol{d}_2^{(1)}$. The explicit expression for the first reflection field $\phi_{12}$ by spheroid $S_2$ is (see equation \eqref{eq:solRodD2})
\begin{multline}    
    \label{eq:phi12Eqn}
    \phi_{12}(\by) = \frac{3}{2a_2^3e_2^3\varepsilon_0}\Bigg[\boldsymbol{d}_2^{(1)}\cdot\bp_2\int_{-a_2e_2}^{a_2e_2} \xi_2\, \mathcal{G}(\by-\bx_2-\xi_2\bp_2)\, d\xi_2 
    \\- \frac{\boldsymbol{d}_2^{(1)}}{2}\cdot(\boldsymbol{\delta}-\bp_2\bp_2)\cdot \boldsymbol{\nabla}_{\by} \int_{-a_2e_2}^{a_2e_2} (a_2^2e_2^2-\xi_2^2)\, \mathcal{G}(\by-\bx_2-\xi_2\bp_2)\, d\xi_2\Bigg].
\end{multline}
% \begin{multline}    
%     \label{eq:phi12Eqn}
%     \phi_{12}(\by) = \frac{3}{2a_2^3e_2^3\varepsilon_0} \boldsymbol{d}_2^{(1)}\cdot\bp_2\int_{-a_2e_2}^{a_2e_2} \xi_2\, \mathcal{G}(\by-\bx_2-\xi_2\bp_2)\, d\xi_2 
%     \\- \frac{3}{4a_2^3e_2^3\varepsilon_0} \boldsymbol{d}_2^{(1)}\cdot(\boldsymbol{\delta}-\bp_2\bp_2)\cdot \boldsymbol{\nabla}_{\by} \int_{-a_2e_2}^{a_2e_2} (a_2^2e_2^2-\xi_2^2)\, \mathcal{G}(\by-\bx_2-\xi_2\bp_2)\, d\xi_2.
% \end{multline}
The dipole moment $\boldsymbol{d}_2^{(1)}$ is given by the Faxén laws as (see equation \eqref{eq:dRod})
\begin{multline}
    \label{eq:d2Eqn}
    \boldsymbol{d}_2^{(1)} = -4\pi a_2^3\frac{3}{2a_2^3e_2^3} \Bigg[ X_2^C \bp_2 \int_{-c_2}^{c_2} \xi_2\, d\xi_2 \int_{a_1e_1}^{a_1e_1}\frac{Q_1}{2a_1e_1} \\
    \, \mathcal{G}(\bx_2+\boldsymbol{\xi}_2 - \bx_1 - \boldsymbol{\xi}_1)\, d\xi_1 
     + \frac{Y_2^C}{2} (\boldsymbol{\delta}-\bp_2\bp_2)\cdot \boldsymbol{\nabla}_{\bx_2} \\
     \int_{-a_2e_2}^{a_2e_2} (a_2^2e_2^2-\xi_2^2) \int_{a_1e_1}^{a_1e_1}\frac{Q_1}{2a_1e_1}\, \mathcal{G}(\bx_2+\boldsymbol{\xi}_2 - \bx_1 - \boldsymbol{\xi}_1)\, d\xi_1   \Bigg],
\end{multline}
where we have used equation \eqref{eq:phi2Eqn} for $\phi_1(\by)$ in place of $\phi^\infty$ in equation \eqref{eq:dRod}. The corresponding first reflection field $\phi_{21}(\by)$ and the dipole moment $\boldsymbol{d}_1^{(1)}$ is obtained by simply switching the labels $1$ and $2$.

The next order correction comes through the second reflection as
\begin{subequations}
    \label{eq:V2Rods}
\begin{gather}
    V_1^{(2)} = \frac{1}{2a_1e_1}\int_{-a_1e_1}^{a_1e_1} \phi_{12}(\bx_1+\xi_1\bp_1)\, d\xi_1, \\
    V_2^{(2)} = \frac{1}{2a_2e_2}\int_{-a_2e_2}^{a_2e_2} \phi_{21}(\bx_2+\xi_1\bp_2)\, d\xi_1,
\end{gather}    
\end{subequations}
with the second reflection fields $\phi_{121}$ and $\phi_{212}$ represented to the leading order in $a/R$ by the dipole moments $\boldsymbol{d}_1^{(2)}$ and $\boldsymbol{d}_2^{(2)}$. These dipole moments can again be obtained using Faxén laws (equation \eqref{eq:dRod}) with first reflection fields in place on $\phi^\infty$. 

Therefore, upto second reflections, the potential on the surface of the spheroids are related to their respective total charges as $V_\alpha = V_\alpha^{(0)}+V_\alpha^{(1)}+V_\alpha^{(2)}$, $\alpha\in \{1,2 \}$. These interactions potentials are accurate upto $\mathcal{O}(a^4/R^4)$.
%Although dipole has direct contributions from G, the charge distribution is such that for large distances the contribution will be from gradG.

\subsection{Far field interaction of a prolate spheroid and a sphere}

Knowing the procedure for two spheroids, it is easy to look at a special case where the second spheroid is a sphere. This simplification is analytically tractable to obtain closed form expressions without losing the flavor of anisotropy in the problem. Consider a spheroid $S_1$ centered at $\bx_1$ with semi-major axis $a$, aspect ratio $\kappa$, eccentricity $e\equiv \sqrt{1-\kappa^{-2}}$ and orientation vector $\bp$, carrying total charge $Q_1$. The second conductor is a sphere $S_2$ centered at $\bx_2$ with radius $\gamma a$ and total charge $Q_2$. The relative separation vector between them is $\bx_{21}\equiv \bx_2-\bx_1\ \equiv -\bx_{12}$.
The relation between the surface potentials of $S_1$ and $S_2$ can be found by either taking limit $e_2\to 0$ in the previous analysis or by applying method of reflection to this system. The results upto the second reflection are stated as follows
\begin{subequations}
    \label{eq:V0RodSphere}
\begin{gather}
    V_1 = \frac{Q_1}{4\pi a\varepsilon_0}\left( \frac{\arctanh e}{e}\right) + V_1^{(1)} + V_1^{(2)}, \\ 
    V_2 = \frac{Q_2}{4\pi\varepsilon_0 \gamma a} + V_2^{(1)} + V_2^{(2)},
\end{gather}
\end{subequations}
where 
\begin{subequations}
\label{eq:V1RodSphere}
\begin{gather}
    V_1^{(1)} = \frac{Q_2}{2ae\varepsilon_0}\int_{-ae}^{ae} \mathcal{G}(\bx_{12}+\xi\bp)\, d\xi, \\
    V_2^{(1)} = \frac{Q_1}{2ae\varepsilon_0}\int_{-ae}^{ae} \mathcal{G}(\bx_{21}-\xi\bp)\, d\xi, 
\end{gather}
\end{subequations}
and
\begin{subequations}
        \label{eq:V2RodSphere}
\begin{gather}
    V_1^{(2)} = -\frac{1}{2ae\varepsilon_0}\int_{-ae}^{ae} \boldsymbol{d}_2^{(1)}\cdot\boldsymbol{\nabla}_{\bx_1}\mathcal{G}(\bx_{12}+\xi\bp)\, d\xi, 
    \\
    V_2^{(2)} = \frac{3}{2a^3e^3\varepsilon_0}\int_{-ae}^{ae} \boldsymbol{d}_1^{(1)}\cdot \Big\{\bp\, \xi\, \mathcal{G}(\bx_{21}-\xi\bp) - \nonumber\\
    \frac{1}{2}(a^2e^2-\xi^2)(\boldsymbol{\delta}-\bp\bp)\cdot\boldsymbol{\nabla}_{\bx_2} \mathcal{G}(\bx_{21}-\xi\bp) \Big\}\, d\xi. 
\end{gather}
\end{subequations}
Here the dipole moments are given by
\begin{subequations}
        \label{eq:d1d2RodSphere}
\begin{gather}    
    \boldsymbol{d}_1^{(1)} = -4\pi a^3 Q_2 \frac{3}{2a^3e^3}\int_{-ae}^{ae} \Big\{ X^C_p\bp\,  \xi\, \mathcal{G}(\bx_{12}+\xi\bp) 
    \nonumber\\
    + \frac{1}{2}Y^C_p  (a^2e^2-\xi^2)(\boldsymbol{\delta}-\bp\bp)\cdot\boldsymbol{\nabla}_{\bx_1} \mathcal{G}(\bx_{12}+\xi\bp) \Big\}\, d\xi ,
    \\
    \boldsymbol{d}_2^{(1)} = -4\pi \gamma^3 a^3 Q_1  \frac{1}{2ae}\boldsymbol{\nabla}_{\bx_2}\int_{-ae}^{ae} \mathcal{G}(\bx_2-\bx_1-\xi\bp)\, d\xi .
\end{gather}
\end{subequations}
These line integrals over $\mathcal{G}$ can be computed analytically \cite{alawneh1977singularity, chwang1975hydromechanics}. After some algebra, we arrive at the closed form expressions for the potentials given by
\begin{subequations}
    \label{eq:V1RodSphere2}
    \begin{gather}
    V_1^{(1)} = \frac{Q_2}{4\pi a \varepsilon_0}\frac{1}{2e}\log\left( \frac{z_{12}-ae-R_-}{z_{12}+ae-R_+} \right),\\  V_2^{(1)} = \frac{Q_1}{Q_2}V_1^{(1)},
\end{gather}
\end{subequations}
where
\begin{gather}
    \label{eq:R1R2App}
    R_\pm \equiv \sqrt{\rho_{12}^2 + (z_{12} \pm ae)^2}, \nonumber \\ \rho_{12}^2 \equiv \bx_{12}\cdot(\boldsymbol{\delta}-\bp\bp)\cdot \bx_{12},
    \quad z_{12} \equiv \bx_{12}\cdot\bp.
\end{gather}
The second order corrections are given by
\begin{multline}
    \label{eq:V2RodSphere2}
    V_1^{(2)} = -\frac{Q_1}{4\pi a\varepsilon_0}\frac{a^2\gamma^3}{4e^2}\Bigg[ \left( \frac{1}{R_-} - \frac{1}{R_+}\right)^2 +  \rho_{12}^2 \\
    \left( \frac{1}{R_+(z_{12}+ae-R_+)} - \frac{1}{R_-(z_{12}-ae-R_-)}  \right)^2 \Bigg], 
\end{multline}

\begin{multline}
    \label{eq:V2RodSphere3}
    V_2^{(2)} = -\frac{Q_2}{4\pi a\varepsilon_0}\frac{9}{4a^2e^6}\Bigg[ 
    X^C_p\Bigg\{ R_- - R_+ \\
    + z_{12}\log\left( \frac{z_{12}-ae-R_-}{z_{12}+ae-R_+} \right) \Bigg\}^2 
    \\
    + \frac{Y^C_p}{4}
    \Bigg\{ \frac{z_{12}}{\rho_{12}}(R_- - R_+) + \frac{ae}{\rho_{12}}(R_- + R_+)  \\- \rho_{12}\log\left( \frac{z_{12}-ae-R_-}{z_{12}+ae-R_+} \right) \Bigg\}^2 \Bigg].
\end{multline}
Recall that $X^C_p$ and $Y^C_p$ are given by equation \eqref{eq:XcYc}.

\subsection{Far field interaction of a oblate spheroid and a sphere}

The eccentricity transformation \eqref{eq:eTransfrm} allows us to directly obtain the surface potential from the prolate spheroid and sphere case, given below. 
\begin{subequations}
    \label{eq:V0DiskSphere}
\begin{gather}
    V_1 = \frac{Q_1}{4\pi a \varepsilon_0}\left(\frac{\kappa \arcsin e}{e}\right) + V_1^{(1)} + V_1^{(2)}, \\
     V_2 = \frac{Q_2}{4\pi\varepsilon_0 \gamma a} + V_2^{(1)} + V_2^{(2)},
\end{gather}
\end{subequations}
The first order corrections are:
\begin{subequations}
    \label{eq:V1DiskSphere}
\begin{gather}
    V_1^{(1)} = \frac{Q_2}{4\pi a \varepsilon_0}\frac{\kappa}{e} \arccot\left( \frac{z_{12}-u}{v-ae/\kappa}\right),\\
    V_2^{(1)} = \frac{Q_1}{Q_2} V_1^{(1)},
\end{gather}
\end{subequations}

where 
\begin{gather}
\label{eq:uvEqnApp}
    u \equiv \sqrt{ \frac{\mu}{2} +
    \sqrt{\frac{\mu^2}{4} + \frac{a^2e^2}{\kappa^2}z_{12}^2} },\, \nonumber\\  
    \mu \equiv |\bx_{12}|^2 - \frac{a^2e^2}{\kappa^2} ,\quad 
    v \equiv \frac{a e z_{12}}{\kappa u}.
\end{gather}
The second order corrections are given by:
\begin{subequations}
    \label{eq:V2DiskSphere}
\begin{gather}    
    V_1^{(2)} = -\frac{Q_1}{4\pi a \varepsilon_0}\frac{\kappa^2a^2\gamma^3}{4 e^2}\Bigg[ \left( \frac{2 v}{u^2+v^2} \right)^2 \nonumber\\
    + \rho_{12}^2\left\{ \frac{4 ae\kappa^{-1} z_{12}-2(z_{12}v+ae\kappa^{-1} u)}{(u^2+v^2)((z_{12}-u)^2+(ae\kappa^{-1}-v)^2)}  \right\}^2 \Bigg],\\
    V_2^{(2)} = -\frac{Q_2}{4\pi a \varepsilon_0}\frac{9 \kappa^6}{a^2 e^6}\Bigg[ 
    X^C_o\left\{ v - z_{12}\arccot\left( \frac{z_{12}-u}{v-ae\kappa^{-1}} \right) \right\}^2 
    \nonumber \\
    + \frac{1}{4}Y^C_o
    \left\{ \frac{ae\kappa^{-1} u - z_{12} v}{\rho_{12}} - \rho_{12}\arccot\left( \frac{z_{12}-u}{v-ae\kappa^{-1}} \right) \right\}^2 \Bigg].
    \end{gather}    
\end{subequations}
Recall that $X^C_o$ and $Y^C_o$ are given by equation \eqref{eq:XcYcDisk}.

\section{Boundary Integral formulation for arbitrary shaped conductors in electrostatics}
\label{app:bim}

The external Dirichlet problem of $N$ charged conductors in an unbounded medium in electrostatics is 
\begin{subequations}
        \label{eq:phiGenEqn}
\begin{gather}
    \nabla^2 \phi(\bx) = 0, \\
    \phi(\bx_s) = V_\alpha, \quad \text{for}\quad \bx_s \in S_\alpha,\\
    \phi(\bx) \to 0 \quad \text{as} \quad |\bx| \to \infty
\end{gather}
\end{subequations}
where $S_\alpha$ denotes surface of conductor $\alpha$ and $\alpha\in \{1,2,...N \}$. 
In the manner similar to micro-hydrodynamics \cite{kim2013microhydrodynamics, pozrikidis1992boundary, pozrikidis2002practical, bagge2023accurate, prosperetti2009computational}, the potential field $\phi(\bx_0)$ can be represented in terms of a double layer potential as 
\begin{multline}
    \label{eq:phiDL}
    \varepsilon_0\phi(\bx_0) = -2\sum_{\alpha=1}^N \oint_{S_\alpha} q_\alpha(\bx) \boldsymbol{\hat n}_\alpha \cdot \boldsymbol{\nabla}_{\bx}\mathcal{G}(\bx,\bx_0)\, dS_\alpha(\bx) \\
    + \sum_{\alpha=1}^N Q_\alpha \mathcal{G}(\bx_0, \bx_\alpha).
\end{multline}
Here the first term denotes the double layer potential, $q_\alpha$ is an unknown double layer density, $\boldsymbol{\hat n}_\alpha$ is outward normal to the surface $S_\alpha$, $Q_\alpha$ is the total charge on $S_\alpha$ and $\bx_\alpha$ is a point lying inside the conductor $S_\alpha$. The unknown double layer densities $q_\alpha$ are determined using the boundary conditions
\begin{equation}
    \label{eq:phiBC1}
    \lim_{\bx_0\to S_\alpha^+} \phi(\bx_0) = V_\alpha, \quad \alpha \in \{1,2,...N \},
\end{equation}
where $\bx_0\to S_\alpha^+$ denotes the approach to the surface $S_\alpha$ is from the outside of the surface, i.e. along $\boldsymbol{\hat n}_\alpha$\footnote{The direction of approach matters because the double layer potential has a jump discontinuity across the surface.}. The second term involving $Q_\alpha$ is needed to complete the double layer representation \cite{kim2013microhydrodynamics, pozrikidis1992boundary}. Applying the boundary condition in equation \eqref{eq:phiDL}, we obtain a second kind integral equations given by
\begin{equation}
    \label{eq:intEqnFull}
    \sum_{\beta=1}^N(\mathcal{L}^d_{\alpha\beta} +  \delta_{\alpha\beta})\, q_\beta(\bx_s) = \sum_{\beta=1}^N Q_\beta \mathcal{G}(\bx_s, \bx_\alpha) - \varepsilon_0V_\alpha,
\end{equation}
where $\alpha \in \{1,2,...N \}$ and $\mathcal{L}^d_{\alpha\beta}$ is the double layer operator given by
\begin{equation}
    \mathcal{L}^d_{\alpha\beta}q_\beta(\bx_s) \equiv 2\oint_{S_\beta} q_\beta(\bx)\, \boldsymbol{\hat n}_\beta \cdot \boldsymbol{\nabla}_{\bx}\mathcal{G}(\bx,\bx_s)\, dS_\beta(\bx), 
\end{equation}
$\bx_s \in S_\alpha.$ 
Given total charges $Q_\alpha$'s on each conductor we are required to obtain the potentials $V_\alpha$'s on the surface of each conductor. Using $\mathcal{L}^d_{\alpha\beta}c = - c\, \delta_{\alpha\beta}$ where $c$ is a constant function defined on the surface of $S_\beta$, we see that \eqref{eq:intEqn} has no unique solution. Since $V_\alpha$'s are unknown, one chooses the solutions $q_\alpha$'s such that the projection of $q_\alpha$ onto the subspace of constant functions (which are eigenfunctions of $\mathcal{L}^d_{\alpha\beta}$) is exactly $V_\alpha$. The corresponding projection operator is given by
\begin{equation}
    \label{eq:Proj}
    \mathcal{P}_{\alpha\beta}^c q_\beta \equiv \frac{1}{|S_\alpha|} \delta_{\alpha\beta}\oint_{S_\beta} q_\beta(\bx)\, dS_\beta(\bx),
\end{equation}
where $|S_\alpha|$ is the surface area of conductor $S_\alpha$. Therefore, choosing $\sum_{\beta=1}^N\mathcal{P}_{\alpha\beta}^c q_\beta = V_\alpha$ not only fixes the non-uniqueness problem but also determines $V_\alpha$'s once the solutions $q_\alpha$'s are known. This leads to a well defined second kind integral equation given by
\begin{equation}
    \label{eq:intEqnFinal}
    \sum_{\beta=1}^N (\mathcal{L}^d_{\alpha\beta} + \mathcal{P}^c_{\alpha\beta} + \delta_{\alpha\beta})\, q_\beta(\bx_s) = \sum_{\beta=1}^N Q_\beta \mathcal{G}(\bx_s, \bx_\alpha), 
\end{equation}
$\alpha \in \{1,2,...N \},$ with the potential fields given by
\begin{equation}
    \label{eq:VBIMFull}
    \varepsilon_0V_\alpha = \frac{1}{|S_\alpha|}\oint_{S_\alpha} q_\alpha(\bx)\, dS_\alpha(\bx),  \quad \alpha \in \{1,2,...N \}.
\end{equation}
Using arguments similar to ones in \cite{kim2013microhydrodynamics, pozrikidis1992boundary}, it can be shown that the spectrum of $\mathcal{L}^d_{\alpha\beta} + \mathcal{P}^c_{\alpha\beta}$ lies in the interval $(-1,1)$ and hence the equation \eqref{eq:intEqnFinal} admits unique solution which can be obtained through Picard iterations. 

To solve the boundary integral equations \eqref{eq:intEqnFinal} for a spheroid and a sphere we perform the surface integrals using Gaussian quadrature defined on the surfaces \cite{golub1969calculation, bagge2023accurate} using the parametric equations of the surfaces. GMRES \cite{bagge2023accurate, saad1986gmres} is used to converge to the solutions.

\bibliography{ref}% Produces the bibliography via BibTeX.

\end{document}